%% file: main.tex
\documentclass[conference]{IEEEtran}

\usepackage{enumitem}
\usepackage{pifont}
\usepackage{multirow,makecell}
\usepackage{tcolorbox}
\usepackage{color}
\usepackage[table]{xcolor}
\usepackage{listings,amsfonts}
\usepackage{threeparttable}
\usepackage{bbding}
\usepackage{graphicx}
\usepackage{tabularx}
\usepackage{booktabs}
\usepackage{longtable}
\usepackage{flushend}
\usepackage{xspace}
\usepackage{tikz}
\usepackage{wasysym}
\usepackage{url}

\AtEndPreamble{
	\usepackage{hyperref}

	\hypersetup{
		colorlinks = true,
		linkcolor = purple,
		anchorcolor = purple,
		citecolor = purple,
		filecolor = purple,
		urlcolor = purple
	}
}

\usepackage{xcolor}
\usepackage{listings}

\definecolor{lstFrame}{RGB}{210,210,210}
\definecolor{lstBg}{RGB}{248,248,248}

\lstdefinestyle{BaseListing}{
  basicstyle=\ttfamily\scriptsize,
  columns=fullflexible,
  keepspaces=true,
  showstringspaces=false,
  breaklines=true,
  breakatwhitespace=false,
  frame=single,
  rulecolor=\color{lstFrame},
  backgroundcolor=\color{lstBg},
  xleftmargin=0.5em,
  xrightmargin=0.5em,
  framexleftmargin=0.4em,
  framexrightmargin=0.4em,
  aboveskip=0.6em,
  belowskip=0.6em,
  captionpos=b,
  numbers=none,
  mathescape=false
}

\definecolor{ghaKey}{RGB}{0,0,150}
\definecolor{ghaString}{RGB}{150,0,0}
\definecolor{ghaComment}{RGB}{0,100,0}
\definecolor{ghaKeyword}{RGB}{0,110,100}

\lstdefinelanguage{GHAYAML}{
  keywords={
    true,false,null,True,False,Null,TRUE,FALSE,NULL,
    yes,no,on,off
  },
  keywordstyle=\color{ghaKeyword}\bfseries,
  sensitive=true,
  comment=[l]{\#},
  commentstyle=\color{ghaComment}\itshape,
  morestring=[b]',
  morestring=[b]",
  stringstyle=\color{ghaString},
  alsoletter={_},
  morekeywords=[2]{
    name,uses,with,run,prompt,env,permissions,jobs,steps,on,
    anthropic_api_key,allowed_non_write_users,claude_args
  },
  keywordstyle=[2]\color{ghaKey}\bfseries
}

\lstdefinestyle{GHAYAMLStyle}{
  style=BaseListing,
  language=GHAYAML
}

\lstdefinestyle{GHAYAMLNumbered}{
  style=GHAYAMLStyle,
  numbers=left,
  numberstyle=\tiny\color{gray},
  stepnumber=1,
  numbersep=6pt,
  firstnumber=1
}

\lstdefinelanguage{WorkflowIR}{
  sensitive=true,
  comment=[l]{\#},
  commentstyle=\color{ghaComment}\itshape,
  alsoletter={_},
  morekeywords={
    WorkflowIR,JobIR,StepIR,ExpressionRef
  },
  keywordstyle=\color{ghaKey}\bfseries,
  morekeywords=[2]{
    str,int,dict,list,Any,None
  },
  keywordstyle=[2]\color{ghaKeyword}\bfseries,
  morestring=[b]',
  morestring=[b]",
  stringstyle=\color{ghaString}
}

\lstdefinestyle{IRSchema}{
  style=BaseListing,
  language=WorkflowIR
}

\newcommand{\promptheading}[1]{\textcolor{ghaKey}{\textbf{#1}}}

\lstdefinestyle{LLMPrompt}{
  style=BaseListing,
  language={},
  keywordstyle=,
  commentstyle=,
  stringstyle=,
  morekeywords={},
  escapeinside={(*@}{@*)}
}

\newcommand{\snipref}[2]{\#Snip.~#1, L#2}

\definecolor{linkblue}{RGB}{0,0,180}
\newcommand{\repo}[1]{\textcolor{linkblue}{\texttt{\nolinkurl{#1}}}}

\newcommand{\xmarkred}{\textcolor{red!75!black}{\ding{55}}}

\newcommand{\tcell}[1]{\makecell[tl]{#1}}
\newcommand{\ncell}[1]{\makecell[tr]{#1}}

\definecolor{best}{HTML}{FFEBEE}   
\definecolor{second}{HTML}{E3F2FD} 

\newcommand{\toolname}{\textsc{TaintAWI}\xspace}

\begin{document}

\title{Demystifying and Detecting Agentic Workflow Injection Vulnerabilities in GitHub Actions}

\author{
\IEEEauthorblockN{
Shenao Wang$^{*,\S}$,
Xinyi Hou$^{*,\S}$,
Zhao Liu$^{\dagger}$,
Yanjie Zhao$^{*}$,
Xiao Cheng$^{\ddagger}$,\\
Quanchen Zou$^{\dagger}$,
Xiangzheng Zhang$^{\dagger}$,
and Haoyu Wang$^{*}$
}
\IEEEauthorblockA{
$^{*}$Huazhong University of Science and Technology, \\
$^{\dagger}$360 AI Security Lab,
$^{\ddagger}$Macquarie University
}
\IEEEauthorblockA{
\{shenaowang, xinyihou, yanjie\_zhao, haoyuwang\}@hust.edu.cn,\\
\{liuzhao3, zouquanchen, zhangxiangzheng\}@360.cn,
xiao.cheng@mq.edu.au
}
}


\maketitle

\begingroup
\renewcommand{\thefootnote}{\S}
\footnotetext{Equal Contribution.}
\endgroup

\input{Chapters/0.abstract}

\IEEEpeerreviewmaketitle

\input{Chapters/1.introduction}
\input{Chapters/2.background}
\input{Chapters/3.motivation}

\input{Chapters/4.preliminary}
\input{Chapters/5.methodology}
\input{Chapters/6.evaluation}
\input{Chapters/7.discussion}
\input{Chapters/8.literature}
\input{Chapters/9.conclusion}

\newpage
\bibliographystyle{IEEEtran}
\bibliography{reference}

\input{Chapters/appendix}

\end{document}

%% file: Chapters/0.abstract.tex
\begin{abstract}
GitHub Actions is increasingly used to deploy LLM-based agents for repository-centric tasks such as issue triage, pull-request review, code modification, and release assistance. These \emph{agentic workflows} extend traditional CI/CD automation with agentic capabilities but also create a new injection surface. In this paper, we introduce \emph{Agentic Workflow Injection}~(AWI), a workflow-level injection flaw where untrusted GitHub event context, such as issue bodies, pull-request descriptions, or comments, is incorporated into agent prompts or agent-consumed inputs and converted into attacker-influenced behavior through agent tools or downstream workflow logic. We identify two core AWI patterns: Prompt-to-Agent~(P2A), where untrusted content reaches an agent prompt boundary, and Prompt-to-Script~(P2S), where attacker influence propagates through model- or agent-derived outputs into later scripts.
We present the first systematic study of AWI in GitHub Actions. We characterize 1,033 real-world AI-assisted actions and extract AWI-specific taint specifications, including prompt boundaries, derived outputs, agentic capabilities, and access-control interfaces. Based on these specifications, we design \toolname{}, a taint-analysis tool that tracks flows from untrusted event context to agent prompt inputs and security-sensitive workflow sinks. Applying \toolname{} to 13,392 real-world agentic workflows from 10,792 repositories, we report 519 potential AWI vulnerabilities, of which 496 are confirmed exploitable under our threat model, yielding a precision of 95.6\%. Among them, 343 are previously unknown zero-day vulnerabilities. We prioritized disclosure for 187 zero-day cases, received 26 maintainer responses, and 24 cases have been accepted or fixed at the time of writing.
\end{abstract}

%% file: Chapters/1.introduction.tex
\section{Introduction}
GitHub Actions~\cite{github_actions} has become a central automation platform for modern software development, where repository maintainers define workflows to build, test, review, triage, and release software through reusable actions and repository-defined scripts~\cite{zheng2026actionfail}. Traditionally, these workflows execute deterministic commands written by maintainers. But this ecosystem is now undergoing a significant shift. An increasing number of workflows now embed Large Language Models~(LLMs) and LLM-based agents through AI-assisted actions such as \texttt{codex-action}, \texttt{claude-code-action}, and \texttt{run-gemini-cli}~\cite{openai_codex_action_2026,anthropic_claude_code_action_2026,google_run_gemini_cli_2026,github_marketplace_ai_assisted_actions_2026}. We refer to such workflows as \emph{agentic workflows}~\cite{github_agentic_workflows_2026}. These \emph{agentic workflows} allow LLM agents to interpret natural-language instructions, reason over repository context, invoke tools, post comments, apply labels, edit files, and interact with GitHub APIs under workflow permissions.

While these agentic capabilities can substantially improve developer productivity, they also broaden the attack surface of CI/CD systems. Recent security reports~\cite{aikido_promptpwnd,snyk_clinejection} have shown that agentic workflows in GitHub Actions can expose privileged agents to attacker-controlled inputs, including issue text, pull-request metadata, commit content, and other collaboration artifacts. Once incorporated into prompts or other agent-facing contexts, such inputs can shape how agents invoke tools, issue commands, and interact with workflow resources. For example, Aikido documented \emph{PromptPwnd} cases in which malicious issues, pull requests, or commit content reached agents with shell-execution or repository-editing capabilities, enabling secret leakage or unauthorized repository actions~\cite{aikido_promptpwnd}. Snyk further analyzed \emph{Clinejection}, where issue-driven prompt injection was chained with GitHub Actions cache poisoning to compromise a downstream release workflow and enable unauthorized package publication~\cite{snyk_clinejection}. Taken together, these cases suggest a broader class of security risks in which untrusted collaboration inputs can affect agentic CI/CD workflows.

To address this emerging risk, we introduce a new type of vulnerability, called \textbf{Agentic Workflow Injection}~(\textbf{AWI}). AWI arises when event contexts controlled by a low-privilege GitHub user, such as issue text, pull-request descriptions, or comments, are exposed to agents in GitHub workflows. We identify two representative AWI patterns. In \textbf{Prompt-to-Agent~(P2A)} AWI, untrusted content reaches an agent prompt boundary and may steer the agent's privileged capabilities. In \textbf{Prompt-to-Script~(P2S)} AWI, attacker influence first passes through a model- or agent-derived output, which is later consumed by downstream workflow logics.
Intuitively, AWI can be viewed as the intersection of \emph{prompt injection}~\cite{liu2025agentfuzz,pedro2025prompt2sql,icse26taintp2x,liu2024llmrce,shi2025promptinjection} and \emph{script injection}~\cite{github_docs_script_injections,github_security_lab_preventing_pwn_requests,openssf_mitigating_attack_vectors_github_workflows,ken_muse_github_actions_injection_attacks} in agentic workflows. However, AWI is not fully captured by either class alone. Unlike conventional prompt injection, AWI targets a CI/CD workflow agent rather than LLM chatbots~\cite{shi2025promptinjection} or traditional LLM-integrated applications~\cite{icse26taintp2x,liu2025agentfuzz,liu2024llmrce}. Unlike traditional script injection, AWI does not require syntactic control over an interpreter. Instead, it relies on semantic influence over an agent or over agent-derived data that later drives privileged workflow operations.

Existing techniques are not sufficient for detecting AWI in agentic GitHub Actions workflows. Prior work on prompt-injection vulnerabilities has studied prompt-to-code~\cite{liu2024llmrce,liu2025agentfuzz}, prompt-to-SQL~\cite{pedro2025prompt2sql}, prompt-to-anything~\cite{icse26taintp2x}, and cross-tool attacks~\cite{wu2026chainfuzzer,li2025crosstoolharvesting}. However, these techniques were not designed for CI/CD workflows and therefore do not capture workflow-specific execution semantics, such as event contexts, action input/output interfaces, and cross-step or cross-job data channels. Therefore, they are difficult to apply directly to AWI detection. Meanwhile, prior CI/CD security research has studied over-privileged workflows~\cite{igibek22characterizing,huang2025revisiting,benedetti22assessment}, GitHub Actions taint analysis~\cite{muralee23argus}, and workflow security scanners~\cite{madjda2026actionscanner}. These tools focus on deterministic workflows and do not model agent prompts, derived outputs, or taint propagation through LLM boundaries. Consequently, we still lack a systematic understanding of how AWI arises in agentic actions and workflows, how prevalent such vulnerabilities are in practice, and how to detect them effectively at the ecosystem scale.

To address this gap, we design \toolname{}, a static analyzer for detecting AWI vulnerabilities in agentic workflows. \toolname{} first characterizes reusable AI-assisted actions by inferring action-level taint specifications, including agentic capabilities, prompt boundaries, derived outputs, and access-control interfaces. It then lifts workflow YAML files into a semantic intermediate representation~(IR), analogous to how a compiler lifts source code into an IR that preserves execution semantics, normalizing raw YAML syntax into structured objects that capture trigger entry points, job-level control flow, and step-level data interfaces. From this IR, \toolname{} constructs an \emph{Agentic Workflow Dependency Graph}~(AWDG) that captures cross-boundary control and data dependency. Finally, \toolname{} performs AWI-oriented taint propagation and reachability analysis to detect P2A and P2S vulnerabilities.


In summary, this paper makes the following contributions:

\begin{itemize}[leftmargin=15pt]
    \item \textbf{AWI Characterization.}
    We introduce Agentic Workflow Injection~(AWI), define its threat model, and characterize two core patterns. We further analyze 1,033 AI-assisted GitHub Actions to identify AWI-relevant attack surfaces that shape downstream workflow risks, finding that only 21 provide explicit caller-identity access control.

    \item \textbf{Practical Detection.}
    We design \toolname{}, a static analyzer for detecting AWI in GitHub Actions workflows. \toolname{} combines action-level taint specifications, workflow dependency modeling, and agent-specific taint propagation to detect P2A and P2S vulnerabilities.

    \item \textbf{Empirical Evaluation.}
    We evaluate \toolname{} on 13,392 real-world agentic workflows from 10,792 repositories. \toolname{} reports 519 potential AWI vulnerabilities, confirms 496 exploitable cases with 95.6\% precision, and identifies 343 previously unknown zero-day vulnerabilities. We prioritized disclosure for 187 zero-day cases, received 26 maintainer responses, and reported 24 cases accepted or fixed at the time of writing.
\end{itemize}

%% file: Chapters/2.background.tex
\section{Background}
\label{sec:background}



\subsection{GitHub Actions Execution Model}

GitHub Actions is an event-driven CI/CD framework in which repository maintainers specify automation logic as \emph{workflows} in YAML files under \texttt{.github/workflows/}~\cite{github_actions}. 

\begin{figure}[t]
    \centering
    \includegraphics[width=\linewidth]{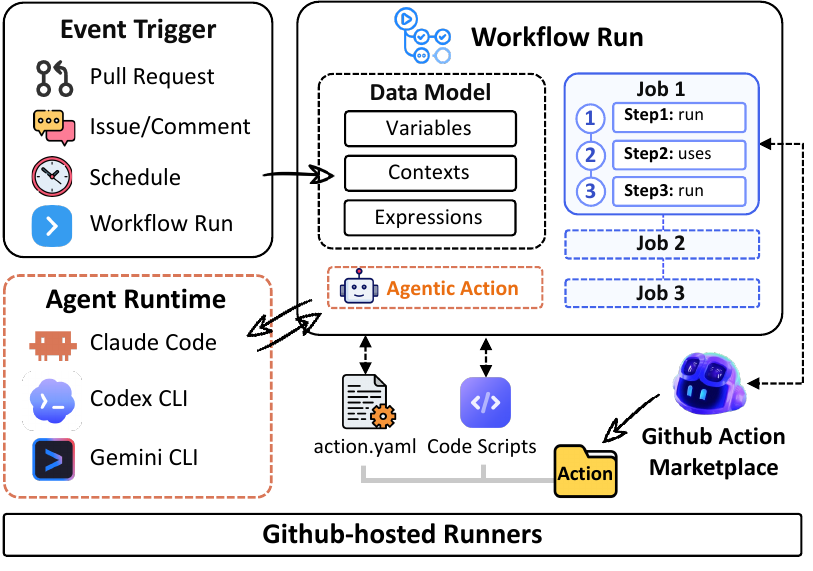}
    \caption{
    GitHub Actions execution model and agentic extensions.
    Solid boxes denote the core workflow components.
    Dashed boxes denote optional workflow elements.
    Orange components highlight agentic workflow extensions.
    }
    \label{fig:gha-agentic-execution}
\end{figure}

\noindent\textbf{Workflow Structure.} As illustrated in \autoref{fig:gha-agentic-execution}, workflows are triggered by repository events, including collaboration events (e.g., \texttt{pull\_request}, \texttt{issues}), scheduled events, manual events (e.g., \texttt{workflow\_dispatch}), and workflow-composition events (e.g., \texttt{workflow\_run})~\cite{github_action_event_trigger}. 
A workflow run instantiates one or more \emph{jobs}, each running on a designated \emph{runner} and consisting of ordered \emph{steps}~\cite{github_actions,github_action_syntax}.
A step either executes commands via \texttt{run} or invokes a reusable \emph{action} via \texttt{uses}~\cite{github_action_syntax}. Actions may be local, imported from external repositories, or obtained from the GitHub Marketplace~\cite{github_marketplace}. 

\noindent\textbf{Data Model.}
GitHub Actions exposes workflow data through \emph{variables}, \emph{contexts}, and \emph{expressions}. \texttt{Variables} store configuration values at scopes such as repository, environment, or workflow scope~\cite{github_docs_variables}. 
\texttt{Contexts} are structured runtime objects that describe the workflow run, triggering event, jobs, and steps~\cite{github_docs_contexts}. Workflow authors access these values using expressions of the form \texttt{\$\{\{ <expression> \}\}}~\cite{github_docs_expressions}, which may appear in conditions, environment assignments, action inputs, and inline scripts.
The \texttt{github} context is security-relevant because it includes the triggering event payload. For example, \texttt{github.event} may expose user-controlled fields such as \texttt{issue.title/body}, \texttt{comment.body}, and \texttt{pull\_request.title/body}~\cite{github_security_lab_untrusted_input}. Once referenced, these values are materialized into workflow fields and consumed by shell scripts, action parameters, environment variables, or downstream jobs. This forms a data-flow channel from externally influenced event payloads to maintainer-defined workflow logic.




\subsection{Agentic Actions and Workflows}

Recent GitHub Actions increasingly integrate LLM agents to automate repository-centric software engineering tasks, such as issue triage, pull-request review, code modification, test debugging, and release assistance~\cite{github_agentic_workflows_2026,github_marketplace_ai_assisted_actions_2026}. We refer to such agent-invoking actions as \emph{agentic actions}, and to workflows that incorporate them as \emph{agentic workflows}. 

\noindent\textbf{Agentic Actions.}
As highlighted in \autoref{fig:gha-agentic-execution}, an \emph{agentic action} is a reusable GitHub Action that invokes an LLM agent within a workflow. Representative examples include \texttt{claude-code-action}~\cite{anthropic_claude_code_action_2026}, \texttt{codex-action}~\cite{openai_codex_action_2026}, and \texttt{run-gemini-cli}~\cite{google_run_gemini_cli_2026}. Some workflows also invoke agent command-line interfaces (CLI) directly through \texttt{run} steps. Although these CLI-based integrations are not actions in the strict terminology, they expose a similar agentic interface and therefore serve the same role.
Agentic actions typically expose a workflow-configurable \emph{prompt interface}, through which maintainers provide tasks or constraints via action inputs, prompt files, environment variables, or command-line arguments. The agent interprets these instructions together with repository state and workflow context and may act through capabilities such as file editing, shell execution, commenting, or labeling~\cite{li2025crosstoolharvesting}.

\noindent\textbf{Agentic Workflows.}
An \emph{agentic workflow} incorporates one or more agentic actions or CLI-based agent steps. It typically constructs an agent-facing task from maintainer-written instructions and runtime workflow data, such as repository files, issue or pull-request content, comments, logs, diffs, and step outputs. Once invoked, the agent executes within the workflow environment and may inherit workspace access, token permissions, and external credentials. As a result, workflow inputs are no longer consumed only by deterministic scripts or action parameters; they may also be interpreted by an agentic layer with access to privileged workflow capabilities. Untrusted repository content can therefore become part of the agent-facing context and influence tool use, generated outputs, or downstream workflow logic under the workflow's authority.

%% file: Chapters/3.motivation.tex
\section{Problem Statement}
\label{sec:awi}
This section defines the problem of AWI in GitHub Actions. We first state our threat model and scope, and then present a real-world motivating example that illustrates how untrusted GitHub event context can be integrated into the agent prompt and converted into attacker-influenced workflow behavior.

\begin{figure*}[t!]
    \centering
    \includegraphics[width=0.9\linewidth]{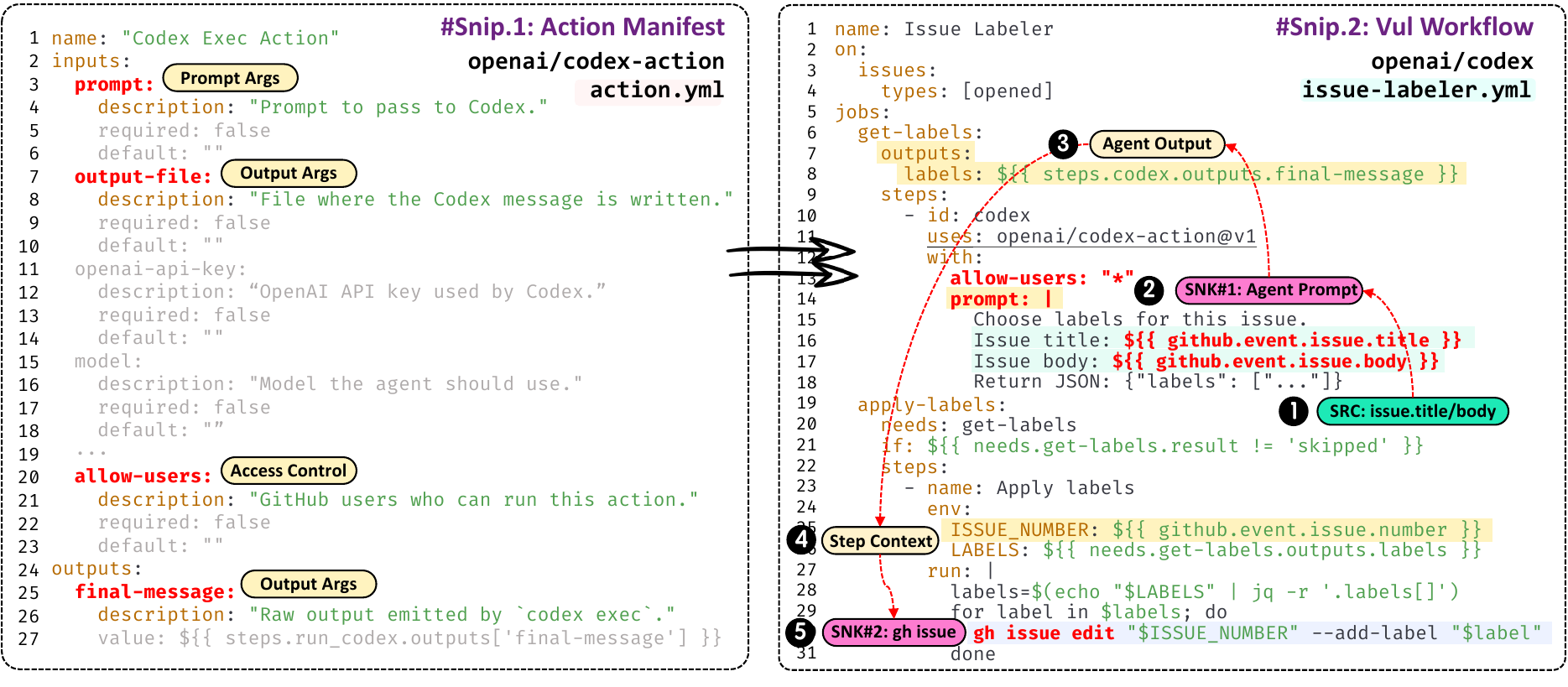}
    \caption{
    A simplified motivating example adapted from \texttt{issue-labeler.yml} in \repo{openai/codex}.
    }
    \label{fig:motivating-example}
\end{figure*}

\subsection{Definition of AWI}

We define \emph{Agentic Workflow Injection} (AWI) as a workflow-level injection flaw where untrusted GitHub event context is incorporated into an agent-facing prompt or other agent-consumed input, and then influences agent behavior or downstream workflow logic under the workflow's authority.
Unlike traditional GitHub Actions script injection, AWI relies on semantic influence rather than syntactically valid code injection. Natural-language content from issues, pull requests, or comments may be interpreted as instructions when mixed with maintainer-written prompts.

\subsection{Threat Model and Scope}
\label{sec:threat-model}

\noindent\textbf{Assumptions.}
We consider GitHub repositories that invoke LLM agents through agentic actions or CLI-based agent steps. These workflows execute within the configured workflow environment and may inherit workspace access, permissions, credentials, and exposed agent capabilities. We do not assume compromise of GitHub, the runner infrastructure, the LLM provider, or maintainer accounts. When an agentic action or CLI enforces sandboxing or tool policies, we treat these mechanisms as part of the configured execution boundary. AWI remains relevant within this boundary because attacker-influenced prompts may cause the agent to exercise its permitted capabilities.

\noindent\textbf{Attacker Capabilities.}
The attacker is an external GitHub user or a low-privilege contributor who can interact with the repository through normal collaboration channels, but cannot modify workflow files, repository settings, protected branches, or secrets. Depending on the workflow trigger, the attacker may create issues, submit pull requests, or post comments. These inputs may appear in GitHub event contexts and be incorporated into agent-facing prompts. The attacker can therefore craft natural-language content that conflicts with maintainer-written instructions and steers the agent, or downstream workflow logic, toward attacker-desired operations such as labeling issues, posting comments, modifying files, executing shell commands, or calling GitHub APIs.

\noindent\textbf{Scope.}
We focus on vulnerabilities caused by unsafe compositions of agentic actions and downstream workflows. In scope are workflows where untrusted repository content is incorporated into an agent-facing prompt, prompt file, command-line argument, environment variable, or workspace file consumed by an LLM agent. We also consider action-level design choices that amplify this risk, such as broad tool capabilities, weak invocation gating for non-write users, and insufficient separation between trusted task instructions and untrusted content. Out of scope are traditional GitHub Actions vulnerabilities without an LLM agent, including script injection~\cite{github_docs_script_injections,hacktricks_gh_actions_context_script_injections}, artifact poisoning~\cite{hacktricks_gh_actions_artifact_poisoning}, and cache poisoning~\cite{hacktricks_gh_actions_cache_poisoning}, as well as attacks requiring runner compromise, sandbox escape, or vulnerabilities in the action implementation itself.

\subsection{Motivating Example}
\label{sec:motivating-example}

\autoref{fig:motivating-example} shows a simplified AWI case adapted from \texttt{issue-labeler.yml} in \repo{openai/codex}. The example combines a reusable agentic action and a downstream issue-labeling workflow. On the action side (\#Snip.1), \repo{openai/codex-action} exposes three interfaces relevant to the AWI path: a configurable \texttt{prompt} input (\snipref{1}{2--6}), an \texttt{allow-users} option for invocation control (\snipref{1}{19--22}), and a \texttt{final-message} output that can be consumed by later workflow steps (\snipref{1}{23--26}). The workflow side (\#Snip.2) shows how these interfaces are composed in a concrete issue-labeling task.

\noindent\textbf{Vulnerable Composition.}
In the \texttt{get-labels} job, the workflow invokes \repo{openai/codex-action} (\snipref{2}{10--11}) and passes issue title and body fields into the action's \texttt{prompt} input (\snipref{2}{14--17}). The action's \texttt{final-message} output is forwarded as the job output \texttt{labels} (\snipref{2}{7--8}). The subsequent \texttt{apply-labels} job consumes this value through \texttt{needs.get-labels.outputs.labels} (\snipref{2}{25}) and runs \texttt{gh issue edit} to apply the labels selected by Codex (\snipref{2}{26--30}). This vulnerable composition exposes two core AWI patterns.

\begin{itemize}
    \item \textbf{Prompt-to-Agent~(P2A).} The first vulnerable path starts from the action input used as an agent prompt. The issue title and body are embedded into the \texttt{prompt} input of \repo{openai/codex-action} (\snipref{2}{14--17}), placing issue-author-controlled content alongside maintainer-written labeling instructions. If the agent interprets such content as instructions rather than task data, the attacker may steer the agent's decisions or tool use.
    \item \textbf{Prompt-to-Script~(P2S).} The second vulnerable path starts from the action output consumed by workflow logic. The action's \texttt{final-message} output is forwarded as the job output \texttt{labels} (\snipref{2}{7--8}), consumed by the \texttt{apply-labels} job through \texttt{LABELS} (\snipref{2}{25--26}), parsed in the shell script, and used to drive \texttt{gh issue edit} (\snipref{2}{28--30}). Here, attacker influence is reflected in the agent-generated output and materialized by conventional workflow logic.
\end{itemize}

%% file: Chapters/4.preliminary.tex
\section{Agentic Action Security Analysis}
\label{sec:action-security}

Agentic actions define how downstream workflows interact with LLM-based agents. In this section, we analyze the action-level properties that shape AWI risks and characterize their presence in real-world AI-assisted GitHub Actions. 
\subsection{Security Properties}
\label{sec:security-properties}

We identify four properties that determine how a reusable action can participate in an AWI path and provide action-specific semantics for downstream workflow analysis.

\noindent\textbf{Prop.\#1: Agentic Capability.}
We first distinguish plain LLM wrappers from agentic actions with operational capabilities. Plain LLM actions typically return text, while agentic actions may read or edit files, execute shell commands, invoke GitHub CLI/API commands, post comments, or apply labels. Such capabilities determine whether prompt influence can be directly materialized inside the action, forming a P2A path.

\noindent\textbf{Prop.\#2: Prompt Configurability.}
AWI requires an agent-facing instruction channel. Reusable actions often expose this channel through action inputs, prompt files, environment variables, command-line arguments, or workspace files. When downstream workflows combine maintainer-written instructions with event-controlled fields, these interfaces become the entry point for untrusted GitHub event context.

\noindent\textbf{Prop.\#3: Access Control.}
Invocation controls determine who can reach the LLM or agent execution path, such as user allowlists, bot allowlists, write-access checks, maintainer approval, or options for non-write users. If a low-privilege user can both supply event content and trigger the action, the workflow exposes a direct attacker-reachable path.

\noindent\textbf{Prop.\#4: Output Exposure.}
Agent-generated results may be exposed to later workflow steps through outputs, generated files, structured JSON, artifacts, comments, or command arguments. When downstream logic treats these values as trusted control data, attacker influence can propagate beyond the agent and be materialized by conventional workflow logic, leading to Prompt-to-Script AWI.

These properties are not vulnerabilities by themselves. They become AWI-relevant when downstream workflows compose them with untrusted event context in ways that expose agent behavior or workflow logic to attacker influence.

\begin{table}[t]
\centering
\caption{Overall characterization of candidate AI-assisted actions.}
\label{tab:action-property-summary}
\setlength{\tabcolsep}{6pt}
\begin{tabular}{p{0.58\linewidth}rr}
\toprule
\textbf{Item} & \textbf{\# Actions} & \textbf{Ratio} \\
\hline
\rowcolor{gray!12}
\multicolumn{3}{l}{\textbf{Action Type}} \\
\quad Agentic & 160 & \cellcolor{red!10}15.5\% \\
\quad LLM-assisted & 572 & \cellcolor{red!25}55.4\% \\
\quad Non-LLM & 301 & \cellcolor{red!14}29.1\% \\
\hline
\rowcolor{gray!12}
\multicolumn{3}{l}{\textbf{AWI-Relevant Interfaces}} \\
\quad Prompt boundary & 453 & \cellcolor{red!20}43.9\% \\
\quad Derived output & 371 & \cellcolor{red!17}35.9\% \\
\quad Access control & 21 & \cellcolor{red!4}2.0\% \\
\hline
\rowcolor{gray!12}
\multicolumn{3}{l}{\textbf{AWI-Relevant Candidates}} \\
\quad Total candidates & 292 & \cellcolor{red!14}28.3\% \\
\quad Agentic candidates & 113 & \cellcolor{red!7}10.9\% \\
\quad LLM-assisted candidates & 179 & \cellcolor{red!9}17.3\% \\
\bottomrule
\end{tabular}
\end{table}

\begin{table*}[t]
\centering
\caption{Widely-adopted agentic and LLM-assisted actions with at least 100 downstream workflow references.}
\label{tab:widely-adopted-action-specs}
\resizebox{\textwidth}{!}{%
\begin{tabular}{llrrlll}
\hline
\textbf{Action} & \textbf{Type} & \textbf{\#Workflow} & \textbf{\#Stars} & \textbf{Prompt Input} & \textbf{Derived Output} & \textbf{Access Control} \\
\hline

\tcell{\repo{anthropics/claude-code-action}}
& \tcell{Agentic}
& \ncell{5,212}
& \ncell{7,366}
& \tcell{\texttt{prompt}}
& \tcell{\texttt{execution\_file}\\\texttt{structured\_output}}
& \tcell{\texttt{allowed\_bots}\\\texttt{allowed\_non\_write\_users}} \\

\tcell{\repo{actions/ai-inference}}
& \tcell{LLM}
& \ncell{795}
& \ncell{471}
& \tcell{\texttt{prompt}, \texttt{prompt-file}\\\texttt{system-prompt}\\\texttt{system-prompt-file}}
& \tcell{\texttt{response}\\\texttt{response-file}}
& \tcell{\xmarkred} \\

\tcell{\repo{openai/codex-action}}
& \tcell{Agentic}
& \ncell{492}
& \ncell{947}
& \tcell{\texttt{prompt}\\\texttt{prompt-file}}
& \tcell{\texttt{final-message}\\\texttt{output-file}}
& \tcell{\texttt{allow-users}\\\texttt{allow-bots}\\\texttt{allow-bot-users}} \\

\tcell{\repo{testdriverai/action}}
& \tcell{Agentic}
& \ncell{300}
& \ncell{5}
& \tcell{\texttt{prompt}}
& \tcell{\texttt{summary}\\\texttt{link}\\\texttt{markdown}}
& \tcell{\xmarkred} \\

\tcell{\repo{google-gha/run-gemini-cli}}
& \tcell{Agentic}
& \ncell{244}
& \ncell{1,963}
& \tcell{\texttt{prompt}}
& \tcell{\texttt{summary}}
& \tcell{\xmarkred} \\

\tcell{\repo{anthropics/cc-base-action}}
& \tcell{Agentic}
& \ncell{227}
& \ncell{804}
& \tcell{\texttt{prompt}, \texttt{prompt\_file}\\\texttt{system\_prompt}\\\texttt{append\_system\_prompt}}
& \tcell{\texttt{execution\_file}\\\texttt{structured\_output}}
& \tcell{\xmarkred} \\

\tcell{\repo{austenstone/copilot-cli}}
& \tcell{Agentic}
& \ncell{195}
& \ncell{17}
& \tcell{\texttt{prompt}}
& \tcell{\texttt{logs-path}\\\texttt{session-path}}
& \tcell{\xmarkred} \\

\tcell{\repo{0xJord4n/aixion}}
& \tcell{LLM}
& \ncell{157}
& \ncell{15}
& \tcell{\texttt{prompt}, \texttt{system}\\\texttt{system\_path}\\\texttt{messages}, \texttt{content}\\\texttt{content\_path}}
& \tcell{\texttt{text}\\\texttt{save\_path}}
& \tcell{\xmarkred} \\

\tcell{\repo{grll/claude-code-action}}
& \tcell{Agentic}
& \ncell{125}
& \ncell{450}
& \tcell{\texttt{custom\_instructions}\\\texttt{direct\_prompt}}
& \tcell{\texttt{execution\_file}}
& \tcell{\xmarkred} \\

\tcell{\repo{mistricky/ccc}}
& \tcell{LLM}
& \ncell{111}
& \ncell{22}
& \tcell{\texttt{from\_tag}\\\texttt{to\_ref}}
& \tcell{\texttt{result}}
& \tcell{\xmarkred} \\

\bottomrule
\end{tabular}
}
\end{table*}

\subsection{Characterizing Agentic Actions in the Wild}
\label{sec:action-characterization}

To understand how these properties appear in practice, we analyze AI-assisted actions in the GitHub Actions Marketplace~\cite{github_marketplace_ai_assisted_actions_2026}. Our goal is not to measure vulnerable workflows yet, but to build an action-level understanding of which reusable actions may expose AWI-relevant interfaces. This characterization provides the action-level semantics that guide our subsequent workflow-level taint analysis.

We collect candidate AI-assisted actions from the \texttt{ai-assisted} category in the GitHub Actions Marketplace~\cite{github_marketplace_ai_assisted_actions_2026} and supplementary keyword-based GitHub searches. After deduplication and relevance filtering, the final set contains 1,035 unique action repositories. We also collect downstream workflows that reuse these actions by searching \texttt{.github/workflows/} files for \texttt{uses: <action>@} references, obtaining 13,392 downstream workflows.
For each candidate repository, we collect action metadata, documentation, package metadata, and source-file context when available. Among the 1,035 candidates, 1,033 contain analyzable action metadata files. We use an LLM-assisted analyzer to infer the four properties in \autoref{sec:security-properties}, including action type, prompt boundaries, derived outputs, and access-control interfaces. The inference focuses on the action metadata interface exposed to workflow authors rather than on providing runtime behavior. The detailed collection and inference procedure is provided in ~\autoref{app:action-collection}. To validate the inference, the first two authors independently review a random sample of 100 inferred records and compare each extracted field against the action metadata, documentation, and examples. The LLM-inferred fields agree with the manual labels in 95.9\% of the reviewed fields. Disagreements are resolved through discussion. For widely adopted actions with over 100 downstream workflow references, we perform a complete manual review to obtain the final action-specific semantics used later.

\subsection{Key Findings}

\noindent\textbf{Overall Findings.}
As shown in \autoref{tab:action-property-summary}, composition-facing interfaces are common, while action-level access controls are rare. Of the 1,033 analyzable actions, 453 expose prompt boundaries and 371 expose model- or agent-derived outputs, but only 21 define explicit caller-identity controls. Among them, 160 actions are agentic and 572 are LLM-assisted.
We treat an agentic action as \emph{AWI-relevant candidate} when it exposes a valid prompt boundary, because attacker-controlled event context can influence the agent execution path. For LLM-assisted actions, we require both a prompt boundary and a derived output, so that model-authored content can reach downstream workflow logic. This leaves 292 AWI-relevant candidates for workflow-level analysis, including 113 agentic actions and 179 LLM-assisted actions.

\noindent\textbf{Widely-Adopted Actions.}
Observed reuse is concentrated in a small number of actions. The most prominent case is \repo{anthropics/claude-code-action}, which appears in 5,212 downstream workflows. We therefore manually review agentic and LLM-assisted actions with at least 100 downstream workflow references. As shown in \autoref{tab:widely-adopted-action-specs}, 10 actions meet this threshold and together account for 7,258 references. These widely adopted actions are dominated by coding-agent integrations, including \repo{anthropics/claude-code-action}, \repo{openai/codex-action}, and \repo{google-gha/run-gemini-cli}. They commonly expose workflow-configurable prompt inputs, such as \texttt{prompt}, \texttt{prompt-file}, \texttt{system\_prompt}, or action-specific instruction fields, and often expose derived outputs, such as \texttt{final-message}, \texttt{summary}, \texttt{structured\_output}, or generated files, that downstream workflow steps can consume.

\noindent\textbf{Access-Control Gap.}
We further find that action-level access control is uncommon and often optional. Among the 10 widely adopted actions in \autoref{tab:widely-adopted-action-specs}, only \repo{anthropics/claude-code-action} and \repo{openai/codex-action} expose explicit caller-identity controls. Even then, protection depends on downstream configuration: a workflow may omit the gate or set it permissively, as in \autoref{fig:motivating-example}, where \texttt{allow-users: "*"} allows non-maintainers to reach the agent execution path.
The remaining eight actions expose no explicit caller gate, so access control must be enforced by workflow logic such as trigger restrictions, permission checks, maintainer approval, or repository-specific guards. Thus, the same reusable action can be safe in a well-gated workflow but attacker-reachable when untrusted GitHub event context flows into its prompt without adequate checks.

%% file: Chapters/5.methodology.tex
\section{Design of \toolname{}}
\label{sec:methodology}

\begin{figure*}[t]
    \centering
    \includegraphics[width=\textwidth]{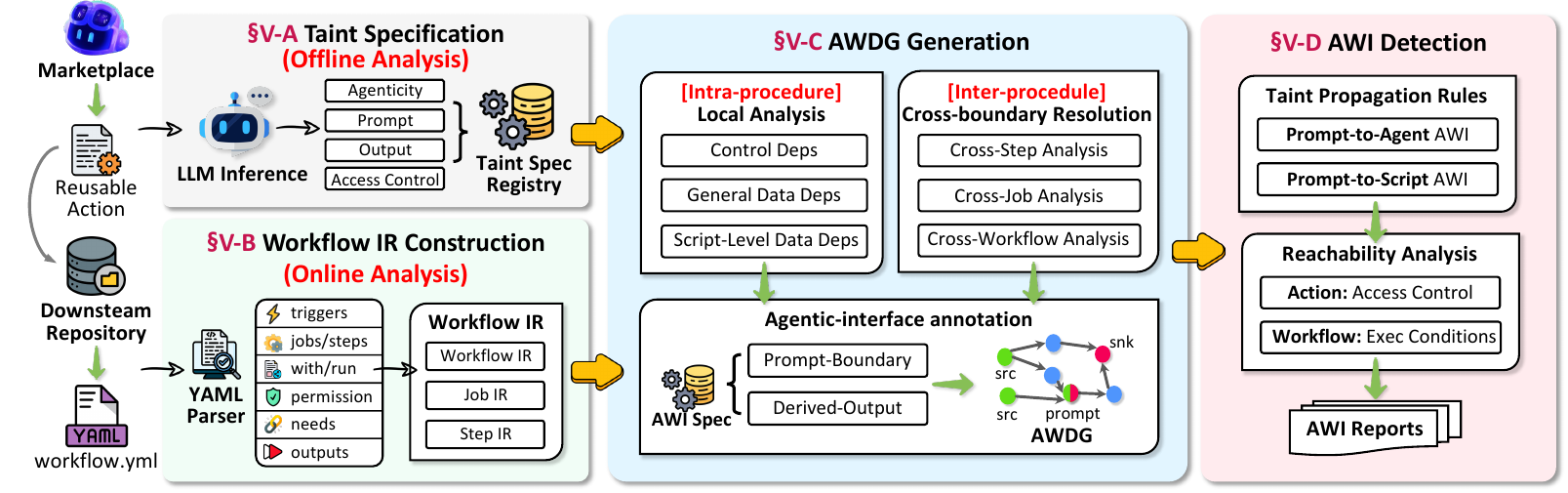}
    \caption{
    Overview of \toolname{} analysis pipeline.
    }
    \label{fig:taintawi-overview}
\end{figure*}

In \autoref{sec:action-security}, we identify the action-level attack surface that shapes AWI risks. Building on those action-level specifications, this section presents \toolname{}, a static analyzer for detecting AWI vulnerabilities in downstream agentic workflows, with an overview shown in \autoref{fig:taintawi-overview}.
\toolname{} first infers action-level taint specifications for reusable AI-assisted actions~(\autoref{sec:taint-spec}), then parses downstream GitHub Actions workflows into a workflow IR~(\autoref{sec:workflow-ir}), constructs an Agentic Workflow Dependency Graph~(AWDG)~(\autoref{sec:awdg}), and performs AWI-specific taint propagation and reachability analysis~(\autoref{sec:taint-propagation}) to report P2A and P2S vulnerabilities.


\subsection{AWI Taint Specification}
\label{sec:taint-spec}

We formulate AWI detection as a taint analysis problem over GitHub Actions workflows. At this stage, \toolname{} defines the security-relevant taint specification. Following the conventional source-sink-sanitizer abstraction, we define an AWI taint specification as:
\[
\mathcal{T} = \langle Src, Snk, San \rangle,
\]
where $Src$ denotes attacker-controllable workflow values, $Snk$ denotes security-relevant locations or operations where attacker influence can materialize, and $San$ denotes sanitizers or guards that block exploitability. \autoref{tab:awi-taint-spec} summarizes representative categories modeled by \toolname{}, with additional examples provided in \autoref{app:awi-taint-spec-details}.

\noindent\textbf{Taint Sources.}
$Src$ contains workflow values that may be influenced by external or low-privilege GitHub users under our threat model. \toolname{} initializes taint from trigger-dependent GitHub event contexts. For collaboration events, such sources include issue, pull-request, comment, review, and discussion fields, such as \texttt{issue.title}, \texttt{issue.body}, \texttt{comment.body}, \texttt{pull\_request.title}, and \texttt{pull\_request.body}. For pull-request and push-related workflows, \toolname{} also models contributor-controlled branch names, commit messages, and related metadata. In addition, serialized event objects, such as \texttt{toJson(github.event)} and \texttt{toJson(github.event.issue)}, are also treated as sources because they may carry multiple attacker-controlled fields into a prompt or script through a single expression. 

\noindent\textbf{Taint Sinks.}
$Snk$ contains locations where attacker influence can materialize as AWI. We distinguish two sink classes according to the two AWI patterns. 

\emph{P2A sinks} are agent-facing prompt interfaces exposed by agentic actions. These include action inputs that are interpreted as natural-language prompts, tasks, instructions, user messages, or system prompts. Such sinks are action-specific and are obtained from the action-level specifications in \autoref{sec:action-security}. For example, \texttt{openai/codex-action.prompt} and \texttt{anthropics/claude-code-action.prompt} are modeled as P2A sinks because tainted data reaching these inputs can directly influence the agent's behavior.

\emph{P2S sinks} are downstream workflow operations that consume model- or agent-derived data and can modify repository, workflow, runner, or external state. Examples include shell steps that invoke \texttt{gh issue edit}, \texttt{gh pr edit}, \texttt{gh api}, or \texttt{git push}. For P2S, \toolname{} also models derived action outputs as semantic taint bridges. When tainted data reaches an agent-facing prompt input, \toolname{} over-approximates the corresponding model- or agent-derived outputs of the same action invocation as attacker-influenced and continues propagation from these outputs. \toolname{} then tracks whether these derived outputs flow into a P2S sink. 

\begin{table}[t]
\centering
\caption{AWI taint specification used by \toolname{}.}
\label{tab:awi-taint-spec}
\setlength{\tabcolsep}{4pt}
\resizebox{1\linewidth}{!}{
\begin{tabular}{c|p{0.34\linewidth}|p{0.6\linewidth}}
\hline
\textbf{Type} & \textbf{Category} & \textbf{Examples} \\
\hline

\multirow{5}{*}{$Src$}
& Issue Context
& \texttt{github.event.issue.body} \\
& Pull-Request Context
& \texttt{github.event.pull\_request.body} \\
& Comment/Review Context
& \texttt{github.event.comment.body} \\
& Branch/Commit Context
& \texttt{github.event.head\_commit.message} \\
& Serialized Event Context
& \texttt{toJson(github.event)} \\

\hline

\multirow{3}{*}{$Snk$}
& P2A Sinks
& \texttt{<agentic-action>.prompt} \\
& P2S Sinks
& \texttt{run: gh issue edit ...}, \\
& 
& \texttt{git push ...} \\

\hline

\multirow{4}{*}{$San$}
& Action-level Guards
& \texttt{allow-users}, \texttt{allowed\_non\_write\_users} \\
& Workflow-level Guards
& \texttt{if: sender == ...}, \texttt{author\_association} \\

\hline
\end{tabular}}
\end{table}

\noindent\textbf{Sanitizers and Guards.}
$San$ contains mechanisms that invalidate an otherwise tainted AWI path. We distinguish action-level guards and workflow-level guards. 

\emph{Action-level guards} are access controls provided by agentic actions, such as \texttt{codex-action.allow-users} and \texttt{claude-code-action.allowed\_non\_write\_users}. These guards restrict which users or bots may cause the agentic action to run. Their effectiveness depends on the downstream workflow configuration; permissive settings such as allowing all users do not sanitize an AWI path.

\emph{Workflow-level guards} are checks implemented by the downstream workflow itself. These include sender allowlists, bot checks, author-association checks, trusted-label gates, maintainer approval, or other \texttt{if} conditions that prevent low-privilege users from reaching the agent execution path. We do not treat prompt-only defenses as sanitizers. Instructions such as ``ignore prompt injection attempts'' remain part of the same natural-language context being attacked and do not enforce a semantic boundary between trusted workflow instructions and untrusted repository content.

\noindent\textbf{AWI Detection Conditions.}
Under this specification, \toolname{} reports a \emph{Prompt-to-Agent}~(P2A) vulnerability when taint from $Src$ reaches a P2A sink without an effective sanitizer, and the target action is agentic. In this case, attacker-controlled event context may be interpreted as agent instructions and converted into tool use within the workflow's execution boundary. \toolname{} reports a \emph{Prompt-to-Script}~(P2S) vulnerability when taint from $Src$ reaches the agent prompt, propagates through a model- or agent-derived output, and then reaches a P2S sink without adequate sanitization. This condition captures workflows that treat agent-generated outputs as trusted control data for privileged downstream operations.

\subsection{Workflow IR Construction}
\label{sec:workflow-ir}

GitHub Actions workflows are written as YAML files, which mix trigger declarations, job permissions, action inputs, inline scripts, environment variables, outputs, and conditional expressions. These fields have different execution semantics and different roles in taint propagation. \toolname{} therefore first lifts each workflow into a workflow intermediate representation~(IR) that preserves the GitHub Actions semantics needed for AWI analysis. As shown in \autoref{lst:workflow-ir-schema}, the IR is organized as a three-level hierarchy of workflow, job, and step objects.

\begin{lstlisting}[
  style=IRSchema,
  caption={Core schema of the workflow IR.},
  label={lst:workflow-ir-schema}
]
WorkflowIR:
  trigger_events: list[str]
  inputs: dict[str, Any]
  env: dict[str, Any]
  permissions: dict[str, Any]
  jobs: list[JobIR]
  expressions: list[ExpressionRef]

JobIR:
  job_id: str
  needs: list[str]
  if_cond: str | None
  env: dict[str, Any]
  permissions: dict[str, Any]
  outputs: dict[str, Any]
  uses: str | None     # reusable workflow call
  with_block: dict[str, Any]
  steps: list[StepIR]

StepIR:
  key: str
  step_id: str | None
  uses: str | None
  run: str | None
  with_block: dict[str, Any]
  env: dict[str, Any]
  if_cond: str | None
\end{lstlisting}

\begin{figure*}[t]
    \centering
    \includegraphics[width=0.97\linewidth]{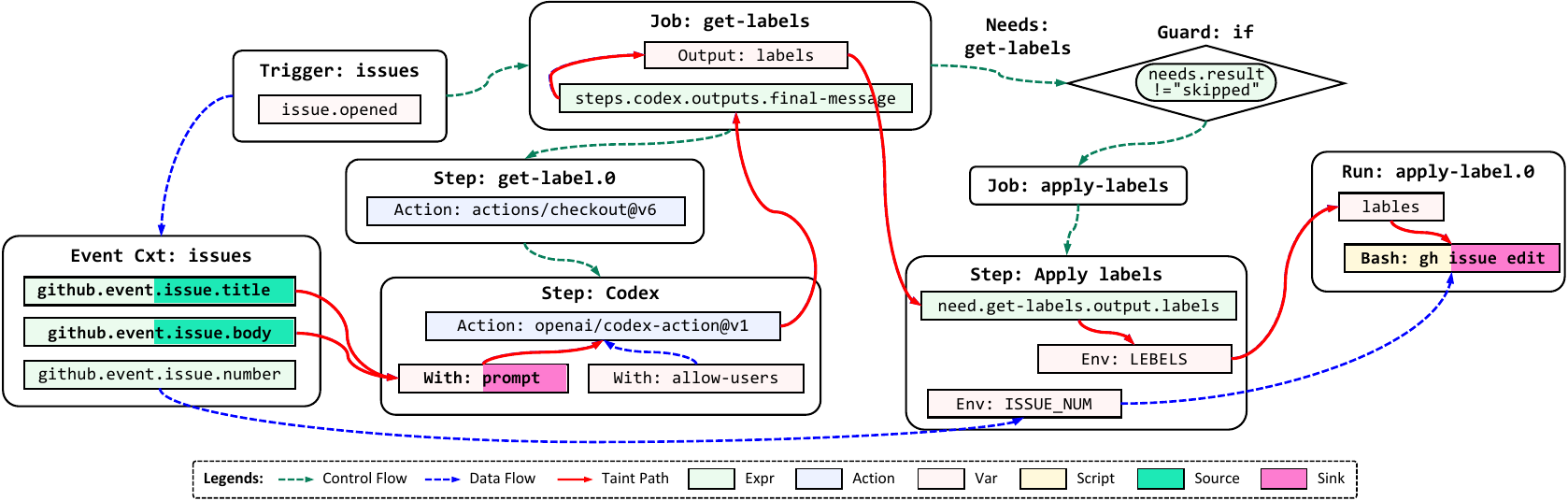}
    \caption{
    Agentic Workflow Dependency Graph~(AWDG) for the motivating example in \autoref{fig:motivating-example}.
    }
    \label{fig:awdg-example}
\end{figure*}

\noindent\textbf{WorkflowIR.}
A \texttt{WorkflowIR} object represents one workflow file. It records the trigger events declared by \texttt{on}, workflow inputs from manual or reusable-workflow interfaces, top-level \texttt{env} and \texttt{permissions}, the list of jobs, and extracted expression references. This level captures the workflow entry points and global configuration. The trigger events determine which GitHub event contexts may become taint sources, while workflow inputs and top-level environment variables define data channels that can be referenced by jobs and steps.

\noindent\textbf{JobIR.}
A \texttt{JobIR} object represents either a normal job or a reusable workflow invocation. It records the job identifier, \texttt{needs} dependencies, job-level \texttt{if} condition, \texttt{env}, \texttt{permissions}, declared job outputs, optional \texttt{uses} target for reusable workflow calls, \texttt{with} inputs, and the ordered list of steps. This level captures job-level execution structure and cross-job data channels. In particular, \texttt{needs} and job outputs provide the interface through which data can flow from one job to another, while job conditions and permissions are later used for reachability and impact reasoning.

\noindent\textbf{StepIR.}
A \texttt{StepIR} object represents an individual workflow step. It records a stable step key, optional step id, \texttt{uses} action, inline \texttt{run} script, \texttt{with} inputs, step-level \texttt{env}, and step-level \texttt{if} condition. This level captures the workflow's main execution sites.  For \texttt{uses} steps, \toolname{} later matches the invoked action against the action taint specification registry to identify prompt inputs, derived outputs, and action-level guards. For \texttt{run} steps, it preserves the script body for later script-level dependency extraction and taint propagation.

The resulting IR normalizes raw workflow YAML into analysis-ready workflow semantics: \texttt{WorkflowIR} captures entry points and global configuration, \texttt{JobIR} captures job-level control and cross-job interfaces, and \texttt{StepIR} captures action invocations and inline scripts.

\subsection{Agentic Workflow Dependency Graph Generation}
\label{sec:awdg}

To capture control and data flows in GitHub Actions workflows, \toolname{} constructs the \emph{Agentic Workflow Dependency Graph}~(AWDG). As illustrated in \autoref{fig:awdg-example}, the AWDG is a directed graph where nodes represent workflow-level values or execution sites, and edges represent potential control flow or data flow. It connects event contexts, expressions, environment variables, action inputs, inline scripts, step outputs, job outputs, files, reusable action invocations, and action-specific prompt/output interfaces. \toolname{} later performs AWI taint propagation over this graph.

Formally, for a workflow $W$, \toolname{} builds an AWDG:
\[
G_W = (V, E_{cf}, E_{df}, \alpha),
\]
where $V$ is the set of graph nodes, $E_{cf} \subseteq V \times V$ is the set of control-flow edges, $E_{df} \subseteq V \times V$ is the set of data-flow edges, and $\alpha$ stores node attributes such as job id, step id, field name, action reference, expression text, and source location. A control-flow edge $(u,v) \in E_{cf}$ indicates that execution may proceed from $u$ to $v$, such as from a workflow trigger to a job, from one step to the next step, or from a producer job to a consumer job through \texttt{needs}. A data-flow edge $(u,v) \in E_{df}$ indicates that the value represented by $u$ may flow into the value or operation represented by $v$, se.g., from an event field to an action input, from an action output to a job output, or from an environment variable to an inline script.

\noindent\textbf{Workflow-local Analysis.}
\toolname{} first constructs control-flow and data-flow edges that can be recovered within a single workflow file. As shown in \autoref{fig:awdg-example}, the workflow-local layer includes the execution structure of the two jobs, their steps, and the fields inside each step. \toolname{} recovers local control flow from triggers, jobs, steps, \texttt{needs}, and \texttt{if} conditions, and connects the corresponding execution nodes with control-flow edges. It then extracts general data-flow edges from GitHub Actions expressions and data-carrying fields. For example, in \autoref{fig:awdg-example}, \texttt{github.event.issue.title} and \texttt{github.event.issue.body} flow into the \texttt{prompt} field of the \texttt{openai/codex-action} invocation through expression interpolation. Similarly, references to \texttt{env}, \texttt{vars}, \texttt{steps}, \texttt{needs}, and \texttt{inputs} are resolved to their corresponding producer nodes whenever possible. For inline \texttt{run} steps, \toolname{} further performs lightweight script-level data-flow analysis to capture common shell flows, including variable assignments, environment reads, command arguments, file writes, and writes to output channels such as \texttt{\$GITHUB\_OUTPUT}. In the example, this analysis connects the \texttt{LABELS} environment variable to the shell variable \texttt{labels} and then to the \texttt{gh issue edit} command.

\noindent\textbf{Cross-boundary Data Resolution.}
\toolname{} then connects flows that cross step, job, or workflow boundaries. In \autoref{fig:awdg-example}, the output of the \texttt{openai/codex-action} step is first exposed as a step output, then promoted to the \texttt{get-labels} job output, and finally consumed by the \texttt{apply-labels} job through \texttt{needs.get-labels.outputs.labels}. To model such paths, \toolname{} resolves step-output references of the form \texttt{steps.<id>.outputs.<name>}, job-output references of the form \texttt{needs.<job>.outputs.<name>}, and caller--callee bindings in reusable workflow calls. These cross-boundary edges make multi-hop propagation explicit, allowing taint to flow from an upstream agent invocation to downstream scripts, actions, or reusable workflows.

\noindent\textbf{Agentic-interface Annotation.}
Finally, \toolname{} enriches the graph with action-specific AWI semantics from \autoref{sec:taint-spec}. For each reusable action invocation, \toolname{} normalizes the action reference and looks up its registered specification. If an input matches a registered prompt boundary, the corresponding node is annotated as an agent-facing prompt boundary. In \autoref{fig:awdg-example}, the \texttt{prompt} input of \texttt{openai/codex-action} is therefore marked as a prompt-boundary node. If an output or file-backed result matches a registered derived output, \toolname{} annotates it as model- or agent-derived; for the motivating example, this marks \texttt{final-message} as an agent-derived output that may carry attacker influence forward. If the action exposes access-control fields, \toolname{} records their configured values as guard metadata on the invocation.

\subsection{Taint Propagation and Vulnerability Detection}
\label{sec:taint-propagation}

Given the AWDG, \toolname{} performs AWI-oriented taint propagation and then checks whether the resulting taint paths are reachable under the workflow and action guards. 

\noindent\textbf{Taint Propagation Rules.}
Let $\tau(v) \in \{\top,\bot\}$ denote the taint state of node $v$, where $\top$ means tainted. Let $Src$ denote attacker-controllable workflow sources, $Prompt(A)$ denote the registered prompt-boundary inputs of an action invocation $A$, and $Output(A)$ denote the registered model- or agent-derived outputs of $A$. We write $u \leadsto v$ if there exists a data-flow path from $u$ to $v$ in the AWDG. \autoref{fig:awi-taint-propagation-rules} summarizes the propagation and detection rules used by \toolname{}.
The first two rules propagate attacker-controlled data through the AWDG. Rule~3 detects Prompt-to-Agent AWI. A prompt input is a P2A sink only when it belongs to an agentic action, because only agentic actions expose operational capabilities such as tool use, shell execution, file modification, or GitHub API operations. Thus, if tainted GitHub event context reaches a registered prompt boundary of an agentic action, \toolname{} reports a P2A path.
Rules~4 and~5 capture Prompt-to-Script AWI. Once a tainted value reaches a registered prompt boundary of an AI-assisted action, \toolname{} treats the registered model- or agent-derived outputs of the same invocation as tainted. This over-approximates the model boundary, since the output may be semantically influenced by the tainted prompt even when there is no syntactic data copy. \toolname{} then continues propagation from these outputs through downstream workflow logic. If a tainted derived output reaches a security-sensitive downstream sink, such as a shell command, GitHub CLI operation, action input, configuration file, or API call, \toolname{} reports a P2S path.

\begin{figure}[t]
\centering
\fbox{
\begin{minipage}{0.95\linewidth}
\footnotesize

\textbf{Notations:}
\begin{itemize}[leftmargin=15pt]
    \item $G_W=(V,E_{cf},E_{df},\alpha)$: AWDG of workflow $W$.
    \item $u,v,w \in V$: AWDG nodes.
    \item $u \leadsto v$: there exists a data-flow path from $u$ to $v$ in $G_W$.
    \item $\tau(u) \in \{\top,\bot\}$: taint state of node $u$.
    \item $Src$: attacker-controllable workflow sources.
    \item $A$: the agentic or AI-assisted action.
    \item $Prompt(A)$: prompt-boundary inputs of action invocation $A$.
    \item $Output(A)$: model- or agent-derived outputs of action invocation $A$.
    \item $Snk_{P2A}$: prompt sinks of agentic actions.
    \item $Snk_{P2S}$: downstream workflow sinks.
\end{itemize}

\textbf{Rules:}
\setlength{\abovedisplayskip}{2.5pt}
\setlength{\belowdisplayskip}{2.5pt}
\begin{enumerate}[leftmargin=15pt, label=\arabic*.]

    \item \textbf{Source Initialization:}
    \[
    u \in Src
    \quad \rightarrow \quad
    \tau(u)=\top
    \]

    \item \textbf{Workflow Data Propagation:}
    \[
    \tau(u)=\top,\; u \leadsto v
    \quad \rightarrow \quad
    \tau(v)=\top
    \]

    \item \textbf{Prompt-to-Agent Detection:}
    \[
    Snk_{P2A} =
    \{u \mid \exists A,\; u \in Prompt(A) \land A \in Agentic\},\;
    \]
    \[
    \tau(u)=\top,\; u \in Snk_{P2A}
    \quad \rightarrow \quad
    \textsf{P2A}
    \]

    \item \textbf{Model/Agent Output Bridging:}
    \[
    \tau(u)=\top,\; u \in Prompt(A),\; v \in Output(A)
    \quad \rightarrow \quad
    \tau(v)=\top
    \]

    \item \textbf{Prompt-to-Script AWI:}
    \[
    \tau(v)=\top,\; v \in Output(A),\; v \leadsto w,\; w \in Snk_{P2S}
    \quad \rightarrow \quad
    \textsf{P2S}
    \]

\end{enumerate}
\end{minipage}
}
\caption{AWI taint propagation and vulnerability detection rules.}
\label{fig:awi-taint-propagation-rules}
\end{figure}

\noindent\textbf{Reachability Analysis.}
Not every taint path is exploitable. After propagation, \toolname{} checks whether the reported path is reachable by an attacker under the workflow's execution conditions and the action's invocation guards. This step filters paths that are present in the AWDG but are unreachable under the threat model in \autoref{sec:threat-model}.

At the action level, \toolname{} examines guard fields recorded in the action specification, such as \texttt{allow-users}, \texttt{allow-bots}, and \texttt{allowed\_non\_write\_users}. A guard is effective only when it prevents the attacker-controlled actor from reaching the action invocation. Permissive configurations, such as wildcard allowlists, are not treated as sanitizers. At the workflow level, \toolname{} examines trigger events, job-level and step-level \texttt{if} conditions, permission settings, and dependency constraints. For example, workflows triggered by \texttt{issues}, \texttt{issue\_comment}, or public pull-request events are considered attacker-reachable unless they contain an effective maintainer-only condition, approval gate, or equivalent access check.

\toolname{} reports a vulnerability only when two conditions hold. First, an AWI taint path exists in the AWDG. Second, the path remains attacker-reachable after applying action-level and workflow-level guards. The final report records the AWI pattern, the source event context, the affected action or downstream sink, the propagation path, and the missing, permissive, or insufficient guards.

%% file: Chapters/6.evaluation.tex
\section{Evaluation}
\label{sec:evaluation}

\begin{table*}[t]
\centering
\scriptsize
\caption{Disclosure status of reported zero-day AWI vulnerabilities with maintainer feedback.}
\label{tab:zero-days}
\setlength{\tabcolsep}{3.5pt}
\resizebox{\textwidth}{!}{%
\begin{tabular}{llrllll}
\toprule
\textbf{Repository} & \textbf{Workflow} & \textbf{\#Stars} & \textbf{Action} & \textbf{Taint Source} & \textbf{Sink Pattern} & \textbf{Status} \\
\midrule

\repo{openai/codex}
& \texttt{issue-labeler.yml}
& 75,373
& \texttt{codex-action}
& Issue title/body
& P2A, P2S
& Acknowledged \\

\repo{dyad-sh/dyad}
& \texttt{claude-triage.yml}
& 20,273
& \texttt{claude-code-base-action}
& Issue title/body
& P2A
& Fixed \\

\repo{OWASP/mastg}
& \texttt{moderator.yml}
& 12,811
& \texttt{ai-inference}
& Issue/PR title/body
& P2S
& Fixed \\

\repo{MIC-DKFZ/nnUNet}
& \texttt{issue-triage.yml}
& 8,289
& \texttt{claude-code-action}
& Issue title/body
& P2A
& Fixed, CVE-2026-44246 \\

\repo{meshtastic/firmware}
& \texttt{models\_issue\_triage.yml}
& 7,189
& \texttt{ai-inference}
& Issue title/body
& P2S
& Accepted \\

\repo{meshtastic/firmware}
& \texttt{models\_pr\_triage.yml}
& 7,189
& \texttt{ai-inference}
& PR title/body
& P2S
& Accepted \\

\repo{InternLM/xtuner}
& \texttt{claude-general.yml}
& 5,112
& \texttt{claude-code-action}
& Comment body
& P2A
& Fixed, CVE-2026-42848 \\

\repo{google-github-actions/run-gemini-cli}
& \texttt{gemini-scheduled-triage.yml}
& 1,974
& \texttt{run-gemini-cli}
& Issue title/body
& P2A
& Acknowledged \\

\repo{Lifailon/lazyjournal}
& \texttt{issue-analysis.yml}
& 1,216
& \texttt{ai-inference}
& Issue title/body
& P2S
& Fixed, CVE Pending \\

\repo{Lifailon/lazyjournal}
& \texttt{pr-check.yml}
& 1,216
& \texttt{ai-inference}
& PR title/body
& P2S
& Fixed, CVE Pending \\

\repo{commons-app/apps-android-commons}
& \texttt{pr\_unrelated\_changes.yml}
& 1,156
& \texttt{ai-inference}
& PR title/body
& P2S
& Fixed \\

\repo{Loukky/gfwlist-by-loukky}
& \texttt{summary.yml}
& 374
& \texttt{ai-inference}
& Issue title/body
& P2S
& Fixed \\

\repo{Devsur11/M5Gotchi}
& \texttt{summary.yml}
& 120
& \texttt{ai-inference}
& Issue title/body
& P2S
& Fixed \\

\repo{ViewTechOrg/Checker-Scammer}
& \texttt{summary.yml}
& 81
& \texttt{ai-inference}
& Issue title/body
& P2S
& Fixed \\

\repo{UnoSite/IsItPayday}
& \texttt{summary.yml}
& 17
& \texttt{ai-inference}
& Issue title/body
& P2S
& Fixed, CVE Pending \\

\repo{RicardoRyn/plotfig}
& \texttt{summary\_new\_issues.yml}
& 9
& \texttt{ai-inference}
& Issue title/body
& P2S
& Fixed \\

\repo{dev-asterix/PgStudio}
& \texttt{summary.yml}
& 6
& \texttt{ai-inference}
& Issue title/body
& P2S
& Accepted, CVE Pending \\

\repo{CI-Till-Krempel/Decide-O-Mat}
& \texttt{gemini-issue-triager.yml}
& 5
& \texttt{run-gemini-cli}
& Issue title/body
& P2A
& Fixed \\

\repo{Exonymos/SeriesScape}
& \texttt{summary.yml}
& 4
& \texttt{ai-inference}
& Issue title/body
& P2S
& Fixed, CVE Pending \\

\repo{flow-pie/chamaa.api}
& \texttt{summary.yml}
& 4
& \texttt{ai-inference}
& Issue title/body
& P2S
& Fixed \\

\repo{HereLiesAz/IDEaz}
& \texttt{jules-issue-handler.yml}
& 3
& \texttt{run-gemini-cli}
& Issue body
& P2A
& Fixed \\

\repo{Leo***Benitez/vision-unlearning}
& \texttt{ai\_issue\_copilot.yml}
& 3
& \texttt{ai-inference}
& Issue title/body
& P2S
& Fixed, CVE Pending \\

\repo{issdandavis/SCBE-AETHERMOORE}
& \texttt{ai-issue-summary.yml}
& 3
& \texttt{ai-inference}
& Issue title/body
& P2S
& Fixed, CVE Pending \\

\repo{OmniBlocks/scratch-gui}
& \texttt{spam-guard.yml}
& 2
& \texttt{ai-inference}
& Comment body
& P2S
& Fixed \\

\repo{OmniBlocks/scratch-gui}
& \texttt{ulti.yaml}
& 2
& \texttt{ai-inference}
& Issue title/body
& P2S
& Fixed \\

\repo{k70***ash1/ai-gen-commit-msg}
& \texttt{auto-commit-message.yml}
& 0
& \texttt{ai-inference}
& PR title/body
& P2S
& Fixed \\

\bottomrule
\end{tabular}
}
\end{table*}

In this section, we evaluate \toolname{} with the following research questions~(RQs):

\noindent\hangindent=2.5em\hangafter=1\textbf{RQ1 [Zero-day]} How many zero-day vulnerabilities can \toolname{} detect but state-of-the-art tools cannot?

\noindent\hangindent=2.5em\hangafter=1
\textbf{RQ2 [Effectiveness]} 
How effective is \toolname{} in detecting AWI vulnerabilities, in terms of detection performance, component contribution, and scalability?

\noindent\hangindent=2.5em\hangafter=1\textbf{RQ3 [Characteristics]} 
What are the characteristics of the detected AWI vulnerabilities, and how do they manifest in real-world workflows?

\subsection{Experimental Setup}

\noindent\textbf{Implementation.}
We implement the prototype of \toolname{} in Python. The implementation follows the design in \autoref{sec:methodology}, including action-level taint specifications, workflow IR construction, AWDG generation, and AWI detection. It also performs lightweight analysis for embedded scripts. For shell \texttt{run} steps, \toolname{} tracks taint through environment variables, local assignments, \texttt{\$GITHUB\_OUTPUT}, \texttt{\$GITHUB\_ENV}, and file-write bridges, followed by rule-based sink refinement. For \texttt{actions/github-script}, it uses YASA~\cite{wang2026yasa} to analyze inline JavaScript after mapping GitHub expressions and \texttt{process.env} references to taint sources. Excluding third-party libraries and evaluation scripts, the online analyzer contains 3.6 KLoC of Python, and the full implementation contains about 4.5 KLoC.

\noindent\textbf{Running Environment.}
All experiments are conducted on a single Ubuntu 22.04.5 LTS server with 256 CPU cores and 1.0~TiB memory. We use Python 3.9.15 and run \toolname{} in a single-machine setting.

\noindent\textbf{Datasets.}
Our evaluation focuses on GitHub Actions workflows that use the AI-assisted or agentic actions characterized in \autoref{sec:action-characterization}. The resulting corpus contains 13,392 workflows that directly invoke at least one agentic or AI-assisted action.

\noindent\textbf{Baselines.}
Since AWI is an emerging vulnerability class, existing detection support remains limited. We compare \toolname{} with two AWI-relevant baseline tools: (1)~\textsc{Prompt Injection Scanner}~\cite{prompt_injection_scanner}~(\textsc{PIScanner} for short), a heuristic workflow scanner that detects unsafe patterns where AI agents process untrusted user input; and (2)~Aikido's \textsc{OpenGrep Rules}~\cite{aikido_opengrep_rules}~(\textsc{OpenGrep} for short), a lightweight pattern-matching baseline for workflow prompt injection. We run both tools with their original rules and settings without modification. 
We also include two workflow taint-analysis baselines, \textsc{Argus}~\cite{muralee23argus} and \textsc{CodeQL}~\cite{codeql_actions_queries}. Since they are general GitHub Actions taint analyzers rather than AWI-specific tools, we supply them with the same agentic-action prompt sink specifications used by \toolname{} so that they can detect P2A source-to-prompt flows. We do not add derived-output specifications or P2S propagation rules to these baselines, because doing so would require implementing the core AWI-specific semantics introduced by \toolname{}. All baseline reports are normalized to workflow-level alerts before comparison.

\subsection{RQ1: Zero-day Vulnerabilities}
\label{sec:rq1-zero-day}

\noindent\textbf{Overall Result.}
\toolname{} initially reported 519 potential AWI vulnerabilities in real-world GitHub Actions workflows. We manually reviewed all reported cases and found that 496 were exploitable under our threat model, while 23 were false positives, corresponding to a precision of 95.6\%. Among the 496 confirmed TP cases, 153 had already been reported by the community but remained unfixed at the time of our analysis. The remaining 343 cases were previously unknown zero-day vulnerabilities.
For responsible disclosure, we prioritized 187 zero-day cases whose repositories either had GitHub stars or had at least one successful workflow-run history, since these cases are more likely to correspond to actively used repositories and functional workflows. Among them, 102 repositories did not provide an available private disclosure channel. For these cases, we avoided publishing technical details and first asked maintainers to add a security policy or provide a private reporting channel. For the remaining 85 cases, we formally reported the vulnerabilities through GitHub Security Advisories, emails, or repository-specific disclosure channels.
At the time of writing, we have received 26 maintainer responses. Among them, 24 cases have been accepted by maintainers or fixed following our reports. \autoref{tab:zero-days} summarizes these fixed or accepted zero-day cases. These fixed or accepted cases affect 23 unique repositories with over 130K GitHub stars in total, and involve multiple AI-assisted actions, including \repo{openai/codex-action}, \repo{anthropics/claude-code-action}, \repo{google-github-actions/run-gemini-cli}, and \repo{actions/ai-inference}.
Representative case studies, including P2S AWI paths in real-world workflows, are provided in \autoref{app:case-studies}.

\begin{table}[t]
\centering
\caption{Comparison with SOTA baselines.}
\label{tab:baseline-comparison}
\begin{threeparttable}
\setlength{\tabcolsep}{5pt}
\begin{tabular}{lrrrrr}
\hline
\textbf{Tool} & \textbf{\#Alerts} & \textbf{\#TP Cov.} & \textbf{\#P2A} & \textbf{\#P2S} & \textbf{\#Missed} \\
\hline
\textsc{PIScanner} 
& 2,079 
& \cellcolor{second}\textbf{78} 
& \cellcolor{second}\textbf{78} 
& 0 
& \cellcolor{second}\textbf{418} \\

\textsc{OpenGrep} 
& 1,658 
& 73 
& 73 
& 0 
& 423 \\


\textsc{Argus} 
& \cellcolor{second}\textbf{831}
& 66 
& 66 
& 0 
& 430 \\

\textsc{CodeQL} 
& 1,486 
& 76 
& 76 
& 0 
& 420 \\

\hline
\toolname{} 
& \cellcolor{best}\textbf{519} 
& \cellcolor{best}\textbf{496} 
& \cellcolor{best}\textbf{103} 
& \cellcolor{best}\textbf{419} 
& \cellcolor{best}\textbf{0} \\
\hline
\end{tabular}
\begin{tablenotes}[flushleft]
\footnotesize
\item \textit{Note.} \#Report denotes raw reports on the full corpus, where fewer reports indicate lower reporting burden. \#TP Cov. denotes covered cases among the 496 confirmed exploitable AWI reports. \#P2A and \#P2S break down covered confirmed cases by pattern. \#Missed denotes confirmed cases missed compared with \toolname{}.
\end{tablenotes}
\end{threeparttable}
\end{table}

\noindent\textbf{Comparison with SOTA.}
\autoref{tab:baseline-comparison} compares \toolname{} with four full-corpus baselines. Since real-world vulnerability detection lacks complete ground truth, we do not report full precision or recall. Instead, we report workflow-level alerts on the full agentic-workflow corpus to reflect reporting burden, and measure coverage over the 496 TP AWI cases confirmed from \toolname{}'s reports.
Overall, \toolname{} reports 519 alerts, among which 496 are confirmed exploitable cases. \textsc{PIScanner} and \textsc{OpenGrep} report 2,079 and 1,658 alerts, respectively, but cover only 78 and 73 confirmed cases. \textsc{PIScanner} heuristically identifies AI-like \texttt{uses:} or \texttt{run} steps using LLM-related keywords and checks whether untrusted GitHub expressions reach their inputs. \textsc{OpenGrep} mainly matches untrusted expressions directly embedded in \texttt{prompt:} fields and selected risky Claude Code configurations. Thus, they can catch some direct source-to-prompt exposures, but do not build workflow-wide dependency graphs or model-derived outputs.
\textsc{Argus} and \textsc{CodeQL} report 831 and 1,486 alerts, and cover 66 and 76 confirmed cases after being supplied with the same agentic-action prompt sink specifications. These taint baselines recover some P2A flows once AI prompt inputs are modeled as sinks, but still stop at the prompt boundary. As a result, they miss P2S cases where such outputs are later consumed by downstream scripts. By modeling both prompt-boundary flows and derived-output propagation, \toolname{} covers 418 additional TPs over the strongest baseline.

\subsection{RQ2: Effectiveness}

\noindent\textbf{FPs \& FNs.}
We evaluate detection performance from both false-positive and false-negative perspectives. For false positives, as reported in \autoref{sec:rq1-zero-day}, \toolname{} reports 519 cases, among which 496 are confirmed exploitable under our threat model, and 23 are false positives, corresponding to a precision of 95.6\%.
Evaluating false negatives is more challenging because real-world GitHub Actions workflows do not have the complete ground truth. We therefore estimate potential false negatives from baseline-only reports, i.e., cases reported by a baseline but not by \toolname{}. Specifically, we randomly sample 100 baseline-only reports from each of \textsc{PIScanner}, \textsc{OpenGrep}, \textsc{Argus}, and \textsc{CodeQL}, resulting in 400 sampled cases in total. After manual review, we identify 13 false negatives of \toolname{}. These missed cases mainly arise from incomplete taint specifications, incomplete script-level sink patterns, and limited script-level propagation semantics. 

\noindent\textbf{Component Contribution.}
We conduct an ablation study over the 496 confirmed TP cases to evaluate the contribution of each major design component in \toolname{}. Specifically, we include four ablation variants.
\emph{w/o Derived-output Bridging} disables the output bridging rule, so taint reaching an agent prompt is not propagated to the corresponding model- or agent-derived outputs.
\emph{w/o Workflow Bridges} removes AWDG edges that connect data across workflow state channels, such as step outputs, job outputs, and reusable-workflow bindings.
\emph{w/o Script Analysis} disables data-flow analysis from embedded shell scripts and \texttt{github-script}.
\emph{w/o Guard Analysis} disables action-level and workflow-level guard filtering, so guarded taint paths may be reported even when they are not attacker-reachable.

\begin{table}[t]
\centering
\caption{Ablation study of \toolname{} over confirmed AWI cases.}
\label{tab:ablation}
\begin{threeparttable}
\setlength{\tabcolsep}{4.5pt}
\begin{tabular}{lrrrr}
\hline
\textbf{Variant} & \textbf{\#Report} & \textbf{\#Covered} & \textbf{\#P2A} & \textbf{\#P2S} \\
\hline
Full \toolname{} 
& 519 
& \cellcolor{best}\textbf{496} 
& \cellcolor{best}\textbf{103} 
& \cellcolor{best}\textbf{419} \\

\emph{w/o Derived-output Bridging} 
& 117 
& 103 
& \cellcolor{best}\textbf{103} 
& 0 \\

\emph{w/o Workflow Bridges} 
& 102 
& 89 
& \cellcolor{second}\textbf{89} 
& 0 \\

\emph{w/o Script Analysis} 
& 462 
& \cellcolor{second}\textbf{442} 
& \cellcolor{second}\textbf{89} 
& \cellcolor{second}\textbf{372} \\

\emph{w/o Guard Analysis} 
& 881 
& \cellcolor{best}\textbf{496} 
& \cellcolor{best}\textbf{103} 
& \cellcolor{best}\textbf{419} \\
\hline
\end{tabular}
\begin{tablenotes}[flushleft]
\footnotesize
\item \textit{Note.} This ablation is evaluated over the confirmed TP AWI cases, not the full workflow corpus. \#Report denotes reports produced within this TP-side ablation run. \#Covered denotes covered cases among the 496 confirmed cases. \#P2A and \#P2S are not mutually exclusive.
\end{tablenotes}
\end{threeparttable}
\end{table}

As shown in \autoref{tab:ablation}, derived-output bridging is essential for P2S detection. Without this component, \toolname{} covers only 103 cases and detects no P2S cases, reducing coverage from 496 to 103. Removing workflow bridges further reduces coverage to 89 cases, showing that many AWI paths rely on workflow data channels. Script analysis also contributes substantially. Disabling it reduces coverage from 496 to 442 and P2S coverage from 419 to 372, indicating that many paths pass through shell scripts or \texttt{github-script} before reaching downstream sinks. Finally, disabling guard analysis increases reports from 519 to 881, suggesting that reachability analysis helps control false positives.

\noindent\textbf{Efficiency.}
We evaluate the efficiency of the online analyzer of \toolname{} on 13,392 agentic workflows. The measured online phase includes workflow IR construction, AWDG generation, taint propagation, and vulnerability detection, but excludes offline action-specification construction, corpus collection, and manual validation.
\toolname{} completes the online analysis in 12.32 minutes on a single server with 256 logical CPU cores and 1008 GB memory. The mean and median analysis times are 0.054 and 0.008 seconds per workflow, respectively. The maximum per-workflow time is 11.360 seconds. Overall, the analyzer achieves a throughput of 18.12 workflows per second. During the run, it reaches a sampled peak resident set size~(RSS) of 0.276 GB and uses 1.11 CPU cores on average. These results show that the online analysis of \toolname{} is practical for large-scale analysis of AI-assisted GitHub Actions workflows.

\subsection{RQ3: Characteristics}
\label{sec:rq3-characteristics}

In this RQ, we characterize AWI from both defensive and vulnerable perspectives. We first examine AWI-relevant taint paths that are blocked by action-level or workflow-level guards, which reflects whether workflow developers have adopted access-control hardening for agentic workflows. We then characterize the 496 confirmed exploitable AWI cases from three aspects: how AWI paths are formed, how attacker influence is materialized, and where vulnerable designs appear in the ecosystem. \autoref{tab:awi-characteristics} summarizes these results. Guarded-path statistics are computed over 881 workflows with AWI-relevant taint paths before reachability analysis, while all other rows are computed over the 496 confirmed exploitable cases. Categories are not mutually exclusive.

\begin{table}[t]
\centering
\caption{Characteristics of AWI-relevant taint paths and confirmed AWI cases.}
\label{tab:awi-characteristics}
\setlength{\tabcolsep}{6pt}
\begin{tabular}{llrr}
\hline
\textbf{Aspect} & \textbf{Category} & \textbf{\#Count} & \textbf{Ratio} \\
\hline

\rowcolor{gray!12}
\multicolumn{4}{l}{\textbf{Guarded Taint Paths}} \\
\multirow{4}{*}{\quad Guard}
& Action-level guard only & 306 & 34.7\% \\
& Workflow-level guard only & 27 & 3.1\% \\
& Both guards & 29 & 3.3\% \\
& Any effective guard & 362 & 41.1\% \\

\hline
\rowcolor{gray!12}
\multicolumn{4}{l}{\textbf{AWI Pattern}} \\
\multirow{4}{*}{\quad Source}
& Issue title/body & 429 & 86.5\% \\
& Pull-request title/body & 50 & 10.1\% \\
& Comment/review body & 24 & 4.8\% \\
& Branch/commit metadata & 18 & 3.6\% \\
\hline
\multirow{3}{*}{\quad Sink}
& Prompt-to-Agent~(P2A) & 103 & 20.8\% \\
& Prompt-to-Script~(P2S) & 419 & 84.5\% \\
& Both P2A and P2S & 26 & 5.2\% \\
\hline
\multirow{3}{*}{\quad Path}
& Short path~($\leq$5 nodes) & 242 & 48.8\% \\
& Medium path~(6--10 nodes) & 205 & 41.3\% \\
& Long path~($>$10 nodes) & 49 & 9.9\% \\
\hline
\rowcolor{gray!12}
\multicolumn{4}{l}{\textbf{Security Impact}} \\
\multirow{4}{*}{\quad Effect}
& GitHub write/API operation & 391 & 78.8\% \\
& Shell/tool execution & 307 & 61.9\% \\
& File/workspace write & 81 & 16.3\% \\
& Network egress & 4 & 0.8\% \\

\hline
\rowcolor{gray!12}
\multicolumn{4}{l}{\textbf{Ecosystem Distribution}} \\
\multirow{7}{*}{\quad Action}
& \texttt{ai-inference} & 393 & 79.2\% \\
& \texttt{run-gemini-cli} & 31 & 6.2\% \\
& \texttt{codex-action} & 14 & 2.8\% \\
& \texttt{claude-code-base-action} & 20 & 4.0\% \\
& \texttt{claude-code-action} & 13 & 2.6\% \\
& \texttt{iflow-cli-action} & 12 & 2.4\% \\
& Other actions & 13 & 2.6\% \\
\hline
\end{tabular}
\end{table}

\noindent\textbf{Guarded Taint Paths.}
We first examine workflows that contain AWI-relevant taint paths but are not reported as vulnerabilities because the paths are blocked by effective guards. Among 881 workflows with AWI-relevant taint paths before guard filtering, 362~(41.1\%) contain an effective guard. Specifically, 306 workflows~(34.7\%) use only action-level guards, such as caller allowlists or non-write-user restrictions; 27~(3.1\%) use only workflow-level guards, such as author-association checks, trusted-label gates, or maintainer-only conditions; and 29~(3.3\%) use both. This result shows that guard placement is highly uneven: action-level guards are much more common than workflow-level guards. In other words, downstream workflows rarely add their own access-control boundary, highlighting the importance of secure-by-default guard mechanisms in reusable agentic actions.

\noindent\textbf{AWI Patterns.}
Among the 496 confirmed exploitable cases, most originate from ordinary collaboration inputs. Issue title/body is the dominant source, appearing in 429 cases~(86.5\%), followed by pull-request fields in 50 cases~(10.1\%) and comment/review bodies in 24 cases~(4.8\%). P2S is the dominant sink pattern, appearing in 419 cases~(84.5\%), while P2A appears in 103 cases~(20.8\%). The path-length distribution further shows that AWI is often not a purely local prompt-construction bug: 205 cases~(41.3\%) have medium-length paths and 49 cases~(9.9\%) have long paths, reflecting multi-hop propagation across GitHub event contexts, agent interfaces, outputs, and downstream workflow logic.

\noindent\textbf{Security Impact.}
The impact of AWI is mainly reflected in state-changing workflow operations. GitHub write/API operations appear in 391 cases~(78.8\%), including issue or PR labeling, commenting, editing, and other repository state changes. Shell or tool execution appears in 307 cases~(61.9\%), and file/workspace writes appear in 81 cases~(16.3\%). These results show that AWI can affect not only model responses, but also CI/CD operations executed with workflow authority.

\noindent\textbf{Ecosystem Distribution.}
The 496 confirmed cases span 440 unique repositories with 421,098 GitHub stars in total, suggesting that AWI-prone designs appear in actively used projects rather than isolated toy examples. They involve multiple AI-assisted actions, with the largest cluster from \texttt{actions/ai-inference} with 393 cases~(79.2\%). This concentration shows how unsafe compositions can emerge when reusable AI actions are combined with untrusted collaboration inputs and downstream automation logic.

%% file: Chapters/7.discussion.tex
\section{Discussion}

\subsection{Security Implications}

\noindent\textbf{For Action Developers.}
Our findings suggest that reusable agentic actions should be treated as security-sensitive workflow primitives. Once an action exposes prompt inputs, tool capabilities, or model-derived outputs, downstream workflows inherit the same trust-boundary problem. However, our action characterization in \autoref{sec:action-characterization} shows that only 21 of 1,033 AI-assisted actions provide explicit caller-identity access control. Even \repo{google-github-actions/run-gemini-cli}, developed by Google, did not originally provide a default caller-gating mechanism, and its official example workflow \texttt{gemini-scheduled-triage.yml} exposed an AWI path before our responsible disclosure. Although this issue has been fixed in the latest version, it suggests that workflow protection is often treated as a downstream user's responsibility.
Unfortunately, our findings in \autoref{sec:rq3-characteristics} reveal that relying on workflow authors to implement guards is fragile. Only 56 of 881 workflows~(6.4\%) with AWI taint paths contain an effective workflow-level guard, whereas 335 workflows~(38.0\%) benefit from action-level guards. Therefore, action developers should implement secure-by-default guard mechanisms in reusable agentic actions. First, agentic actions should provide built-in access control, such as write-permission checks, caller allowlists, or explicit opt-in for non-write users. Second, actions should separate trusted instructions from untrusted repository content in their interfaces. Third, if possible, actions should expose typed or constrained outputs, such as validated label sets, instead of free-form model responses. These mechanisms have been well adopted by \texttt{claude-code-action}~\cite{anthropic_claude_code_action_security} and \texttt{codex-action}~\cite{openai_codex_action_security}.

\noindent\textbf{For Workflow Developers.}
Workflow developers should treat GitHub event context as untrusted whenever it can be influenced by external users. Many vulnerable workflows directly rely on prompt-level protection, such as asking the agent to ignore prompt injection. Such soft constraints are insufficient because they are interpreted by the same model that receives the adversarial content. Developers should instead enforce workflow-level access control, such as actor-permission checks, maintainer-applied labels, approval gates, trusted trigger conditions, and least-privilege permissions.
For public workflows such as issue triage, developers should avoid directly interpolating \texttt{github.event.issue.title} or \texttt{github.event.issue.body} into maintainer-written prompts. A safer pattern is to pass only a stable identifier, such as \texttt{github.event.issue.number}, and let the agent retrieve the issue content through constrained commands such as \texttt{gh issue view}. In addition, model-derived outputs should be validated with strict allowlists or typed schemas before being consumed by downstream workflows.

\subsection{Limitations}

\noindent\textbf{Static Over-approximation.}
\toolname{} is a static analyzer and therefore over-approximates possible workflow executions. It may report paths that are syntactically feasible but difficult to exploit in practice due to repository-specific policies, unavailable secrets, failed workflow runs, or undocumented runtime behavior of third-party actions. We mitigate this issue with reachability analysis and manual validation for reported cases, but fully modeling GitHub's runtime behavior and each action's internal implementation remains challenging.

\noindent\textbf{Incomplete Action Semantics.}
\toolname{} relies on action-level taint specifications, including prompt inputs, derived outputs, access-control options, and agentic capabilities. In this work, we focus on popular agentic actions that appear in real-world workflows. However, the GitHub Actions ecosystem evolves quickly, and long-tail or newly released actions may expose prompt or output interfaces that are not yet modeled. Missing or outdated specifications can lead to false negatives.

\noindent\textbf{Limited Embedded-script Analysis.}
GitHub workflows can embed or invoke scripts written in multiple languages. \toolname{} currently performs lightweight analysis for common shell \texttt{run} steps and uses YASA to analyze inline JavaScript in \texttt{actions/github-script}. This design is sufficient to recover many practical AWI paths, but it does not provide complete data flow analysis. In particular, complex shell semantics, cross-file scripts, and dynamically generated commands may be missed. We view more complete multi-language script analysis as important future work.

%% file: Chapters/8.literature.tex
\section{Related Work}
\label{sec:related}

\noindent\textbf{Security of GitHub Actions and CI/CD Workflow.}
Existing work on GitHub Actions security has focused on workflow misconfiguration, injection risks, and automated analysis. Koishybayev et al.~\cite{igibek22characterizing} characterized the security properties of GitHub CI workflows and showed that over-privileged permissions, untrusted triggers, and third-party action dependencies are widespread in practice. Benedetti et al.~\cite{benedetti22assessment} demonstrated the prevalence of workflow security weaknesses through automated assessment, while Muralee et al.~\cite{muralee23argus} introduced ARGUS, a staged static taint analysis framework for detecting code-injection vulnerabilities in GitHub workflows and actions. Recent work revisited workflow security practices over time~\cite{huang2025revisiting}, compared the scope and effectiveness of workflow scanners~\cite{madjda2026actionscanner}, and analyzed failed workflow executions to build a taxonomy of failure causes~\cite{zheng2026actionfail}. Practitioner guidance from GitHub Security Lab, GitHub Docs, OpenSSF, and Ken Muse documented concrete attack vectors such as pwn requests, untrusted input expansion, and shell-script injection~\cite{github_security_lab_preventing_pwn_requests,github_security_lab_untrusted_input,github_docs_script_injections,openssf_mitigating_attack_vectors_github_workflows,ken_muse_github_actions_injection_attacks}. Most of this literature studies traditional, deterministic workflows. Recent incident reports such as PromptPwnd and Clinejection show that introducing agentic capabilities can expose new vulnerabilities in GitHub workflows~\cite{aikido_promptpwnd,snyk_clinejection}. However, these reports are case-driven and do not provide a systematic understanding of the real-world prevalence and impact of this vulnerability class.

\noindent\textbf{Emerging Attack Surfaces in LLM-Integrated Systems.}
LLM integration allows untrusted natural-language input to influence tool calls, code execution, and other external actions~\cite{weber2024llmapp,wang2024agent,he2024emerged,deng2025agents}. Empirical studies have identified prompt-to-code execution~\cite{liu2024llmrce}, prompt-to-SQL injection~\cite{pedro2025prompt2sql}, prompt-to-anything vulnerabilities~\cite{icse26taintp2x}, and broader taint-style attacks in tool-using agents~\cite{liu2025agentfuzz}. Follow-up work further studied prompt injection in tool selection~\cite{shi2025promptinjection}, cross-tool harvesting and pollution~\cite{li2025crosstoolharvesting}, and multi-tool vulnerabilities in LLM agents~\cite{wu2026chainfuzzer}, while benchmarks and auditing frameworks began to systematize agent security evaluation~\cite{zhang2024agent,luo2026agentauditor}. Emerging auditing and scanning tools also reflect growing practical concern about this class of risks~\cite{trailofbits_agentic_actions_auditor,prompt_injection_scanner,aikido_opengrep_rules}. However, existing studies primarily examine LLM-integrated applications and agents in general settings, leaving their interaction with CI/CD workflow semantics underexplored. Our work fills this gap by systematically studying how LLM capabilities in real-world GitHub Actions workflows enable untrusted inputs to affect downstream workflow behavior.

%% file: Chapters/9.conclusion.tex
\section{Conclusion}
\label{sec:conclusion}

This paper introduced Agentic Workflow Injection~(AWI), a workflow-level injection vulnerability in agentic GitHub Actions workflows. We identified two core AWI patterns and designed \toolname{}. We characterized 1,033 real-world AI-assisted actions and found that many expose prompt boundaries or model-derived outputs, while few provide caller-identity access control. On 13,392 agentic workflows from 10,792 repositories, \toolname{} reported 519 potential AWI vulnerabilities, of which 496 were confirmed exploitable under our threat model, including 343 previously unknown zero-day vulnerabilities. Our findings show that AWI is a practical risk arising from reusable agentic actions, highlighting the need for secure-by-default action interfaces and safer workflow composition practices.

%% file: Chapters/appendix.tex
\section{Appendix}

\subsection{Dataset Collection and Property Inference}
\label{app:action-collection}

This appendix details the dataset construction and property inference process used in our characterization and evaluation. We collect candidate AI-assisted actions, gather downstream workflows that invoke them, and infer security-relevant action properties for subsequent taint analysis.

\textbf{Candidate Action and Workflow Collection.}
We collect candidate actions with potential agentic or LLM-based behavior from two sources. We crawl the \texttt{ai-assisted} category in the GitHub Actions Marketplace~\cite{github_marketplace_ai_assisted_actions_2026}, obtaining 1,027 marketplace records. To cover emerging actions outside the Marketplace, the first two authors also conduct independent keyword-based GitHub searches using terms such as LLM, AI agent, and GitHub Action. After deduplication and relevance filtering, this adds 8 actions, including \repo{openai/codex-action} and \repo{xiaoju111a/kimi-actions}, resulting in 1,035 unique action repositories.
For each resolved candidate action, we search workflow files under \texttt{.github/workflows/} that invoke the action through references of the form \texttt{uses: <action>@}. We search both \texttt{.yml} and \texttt{.yaml} files, exclude forked repositories, and recursively split oversized queries by file-size ranges to reduce GitHub code-search truncation. This process yields 14,223 search hits, corresponding to 14,174 unique downstream workflow files from 10,851 repositories. After resolving repository snapshots, our final corpus contains 78,477 workflow files from 10,792 repositories, covering 230,723 action invocations. Among them, 13,392 workflows directly invoke at least one queried AI-assisted action and serve as the primary workflow-level evaluation set.


\textbf{Security Property Inference.}
For each candidate repository, we collect the artifacts that describe how the action is configured and used, including \texttt{action.yml} or \texttt{action.yaml}, README files, \texttt{package.json}, and repository files when available. Among the 1,035 candidate repositories, 1,033 contain analyzable action metadata files and are included in the inference step. We then use an LLM-assisted analyzer to characterize the four security-relevant properties defined above. \autoref{lst:action-property-prompt} shows the simplified prompt. The LLM is given action metadata, documentation excerpts, package metadata, and source-file context, and is required to return a fixed-schema record.
The inference focuses on the action metadata interface exposed to workflow authors.
We classify an action as \texttt{agent} only when the LLM is given an agent-style control loop or tool calling capabilities. We classify an action as \texttt{llm} when it exposes plain inference, chat, summarization, or review behavior without clear agentic tool use. For prompt interface, we record only inputs or documented environment variables whose values directly become natural-language prompts, task descriptions, user messages, or system prompts. For derived output interface, we record only model- or agent-authored values that can be consumed by later workflow steps. For access control interface, we record only caller-identity gates, such as user allowlists, bot allowlists, non-write-user controls, or write-access checks.

\begin{lstlisting}[
  style=LLMPrompt,
  float=t,
  caption={Simplified prompt used for security property inference.},
  label={lst:action-property-prompt}
]
(*@\promptheading{[Task]}@*)
Analyze the GitHub Action repository and infer
security-relevant action properties. This is a semantic
classification task, not a vulnerability detection task.

(*@\promptheading{[Input Context]}@*)
Use only evidence from:
- action.yml/action.yaml
- README
- package metadata
- source files

(*@\promptheading{[Output Schema]}@*)
Return exactly one JSON object with these fields:
1) action: exact action name;
2) is_agentic: one of agent, llm, or no;
3) prompt_boundaries: model-readable prompt or instruction interfaces;
4) derived_outputs: model-authored outputs or files consumed downstream;
5) access_control: caller-identity access-control interfaces.

(*@\promptheading{[Decision Rules]}@*)
- prompt_boundaries include only inputs that become
  natural-language prompts, tasks, or instructions.
- derived_outputs include only model- or agent-authored
  values that can be consumed by later workflow steps.
- access_control includes only caller allowlists,
  write-access gates, or equivalent identity checks.
- If evidence is uncertain, leave the field empty.
\end{lstlisting}

\textbf{Manual Validation.} We validate the LLM-assisted inference results through manual review to ensure the accuracy of the extracted properties. The first two authors independently review a random sample of 100 inferred records and compare each extracted field against the action metadata, documentation, and examples. The LLM-inferred fields agree with the manual labels in 95.9\% of the reviewed fields. Disagreements are resolved through discussion and used to refine the extraction criteria. For widely-adopted actions with over 100 downstream workflow references~(22 of 1033), we perform a complete manual review to obtain the final action-specific semantics used in later taint analysis.

\subsection{AWI Taint Specification Patterns}
\label{app:awi-taint-spec-details}

\autoref{tab:awi-taint-spec-details} expands the AWI taint specification in \autoref{tab:awi-taint-spec} with representative sources, sinks, and sanitizers used by \toolname{} to model attacker-controlled workflow context, AI-facing prompt interfaces, downstream workflow operations, and access-control guards.

\begin{table*}[t]
\centering
\caption{AWI taint specification used by \toolname{}.}
\label{tab:awi-taint-spec-details}
\setlength{\tabcolsep}{5pt}
\resizebox{0.9\linewidth}{!}{
\begin{tabular}{c|c|l}
\hline
\textbf{Type} & \textbf{Category} & \textbf{Representative Examples} \\
\hline

\multirow{5}{*}{$Src$}
& Issue Context
& \texttt{github.event.issue.title}, \texttt{github.event.issue.body} \\
& Pull-Request Context
& \texttt{github.event.pull\_request.title}, \texttt{github.event.pull\_request.body} \\
& Comment/Review Context
& \texttt{github.event.comment.body}, \texttt{github.event.review.body} \\
& Branch/Commit Context
& \texttt{github.event.pull\_request.head.ref}, \texttt{github.event.head\_commit.message} \\
& Serialized Event Context
& \texttt{toJson(github.event)}, \texttt{toJson(github.event.issue)} \\

\hline

\multirow{5}{*}{$Snk$}
& P2A Sinks
& \makecell[l]{
\texttt{anthropics/claude-code-action.prompt}\\
\texttt{openai/codex-action.prompt}\\
\texttt{google-github-actions/run-gemini-cli.prompt}
} \\
\cline{2-3}
& P2S Sinks
& \makecell[l]{
\texttt{run: gh issue edit ...}\\
\texttt{run: gh pr edit ...}\\
\texttt{run: gh api ...}\\
\texttt{run: git push ...}
} \\

\hline

\multirow{4}{*}{$San$}
& Action-level Guards
& \makecell[l]{
\texttt{anthropics/claude-code-action.allowed\_non\_write\_users}\\
\texttt{anthropics/claude-code-action.allowed\_bots}\\
\texttt{openai/codex-action.allow-users}\\
\texttt{openai/codex-action.allow-bots}
} \\
\cline{2-3}
& Workflow-level Guards
& \makecell[l]{
\texttt{if: github.event.sender.login == ...}\\
\texttt{if: github.event.sender.type == 'Bot'}\\
\texttt{if: github.event.pull\_request.author\_association == 'COLLABORATOR'}
} \\

\hline
\end{tabular}}
\end{table*}

\subsection{Case Studies}
\label{app:case-studies}
This appendix provides representative case studies for the AWI vulnerabilities identified in \autoref{sec:rq1-zero-day}. They illustrate how attacker-controlled issue content can enter AI prompts, influence model-generated outputs, and later reach privileged workflow logic, leading to Prompt-to-Script AWI in real-world GitHub Actions workflows.

\noindent\textbf{Case\#1: Prompt-to-Script AWI in run-gemini-cli.} \autoref{lst:gemini-triage-case} shows a simplified snippet adapted from the example workflow \texttt{gemini-scheduled-triage.yml} in \repo{google-github-actions/run-gemini-cli}. The workflow collects attacker-controlled issue metadata and exposes it through \texttt{issues\_to\_triage} (L6--L8), then injects this value into Gemini's prompt via \texttt{echo \$ISSUES\_TO\_TRIAGE} (L15--L18). Gemini's structured output is forwarded to the \texttt{apply-labels} job through \texttt{needs.triage.outputs.triaged\_issues} (L27--L28) and used in \texttt{github.rest.issues.setLabels} (L31--L35). This forms a Prompt-to-Script AWI path because the downstream script trusts the model-produced \texttt{issue\_number} and \texttt{labels\_to\_set} under write permissions. Although the workflow validates labels, it does not verify that \texttt{issue\_number} belongs to the originally triaged issues, allowing an injected issue to relabel an unintended issue or PR.

\noindent\textbf{Implications.}
At the time of writing, the affected repository
\repo{google-github-actions/run-gemini-cli} has 1,974 stars, and the vulnerable workflow is an official example provided by the repository. As a result, the pattern can be copied by downstream users. Based on our findings, more than 100 downstream workflows adopted this example or closely
related variants. We responsibly disclosed this issue through Google AI VRP, and Google updated the affected workflow in PR\#505. In the same report, we also discussed the need for secure-by-default action-level access controls. Google subsequently updated its trust and security guidelines to include this concern.

\begin{lstlisting}[
  caption={Prompt-to-Script AWI in \texttt{gemini-scheduled-triage.yml}.},
  label={lst:gemini-triage-case},
  style=GHAYAMLNumbered,
  numbers=left,
  numberstyle=\tiny,
  xleftmargin=1.8em,
  framexleftmargin=1.2em
]
jobs:
  triage:
    outputs:
      triaged: ${{ steps.gemini.outputs.summary }}
    steps:
      - id: find_issues
        run: |
          gh issue list --json number,title,body > issues.json
          echo "issues=$(cat issues.json)" >> "$GITHUB_OUTPUT"

      - id: gemini
        uses: google-github-actions/run-gemini-cli@v0
        env:
          ISSUES: ${{ steps.find_issues.outputs.issues }}
        with:
          prompt: |
            Issues to triage:
            !`{echo $ISSUES}`

            Return JSON with issue_number and labels_to_set.

  apply-labels:
    needs: triage
    steps:
      - uses: actions/github-script@v8
        env:
          TRIAGED: ${{ needs.triage.outputs.triaged }}
        with:
          script: |
            for (const issue of JSON.parse(process.env.TRIAGED)) {
              await github.rest.issues.setLabels({
                issue_number: issue.issue_number,
                labels: issue.labels_to_set,
                ...
              });
            }
\end{lstlisting}

\noindent\textbf{Case\#2: Prompt-to-Script AWI resulting in command execution in GitHub's \texttt{summary.yml}.} \autoref{lst:summary-shell-case} shows \texttt{summary.yml} in the \repo{actions/starter-workflows}, GitHub's official workflow template repository. A public event (\texttt{issues.opened}) supplies attacker-controlled issue title and body to the prompt of the \texttt{actions/ai-inference} action (L23--L26), and the generated response is then interpolated into \texttt{gh issue comment \$ISSUE\_NUMBER --body '\${{ steps.inference.outputs.response }}'} (L30). Because GitHub expression interpolation occurs before Bash parses the script, a model-generated single quote can terminate the intended \texttt{--body} string and cause the remainder of the response to be interpreted as shell syntax. With \texttt{issues: write} permission and \texttt{GH\_TOKEN} exposed to the shell step, this creates a P2S command-execution sink from untrusted issue content to privileged shell execution.

\noindent\textbf{Implications.}
At the time of writing, the official template repository \repo{actions/starter-workflows} has 11,535 GitHub stars. Although GitHub updated \texttt{summary.yml} in late February 2026 to replace direct interpolation with \texttt{gh issue comment \$ISSUE\_NUMBER --body "\$RESPONSE"}. However, among the 519 AWI findings reported by \toolname{}, 93 downstream workflows reused the older pre-fix version of \texttt{summary.yml} and inherited the same vulnerable prompt-to-shell path. 
We responsibly disclosed these vulnerabilities to the affected repositories and recommended updating to the patched template.

\begin{lstlisting}[
  caption={Prompt-to-Script AWI resulting in command execution in \texttt{summary.yml}.},
  label={lst:summary-shell-case},
  style=GHAYAMLNumbered,
  numbers=left,
  numberstyle=\tiny,
  xleftmargin=1.8em,
  framexleftmargin=1.2em
]
name: Summarize new issues

on:
  issues:
    types: [opened]

jobs:
  summary:
    runs-on: ubuntu-latest
    permissions:
      issues: write
      models: read
      contents: read

    steps:
      - name: Checkout repository
        uses: actions/checkout@v4

      - name: Run AI inference
        id: inference
        uses: actions/ai-inference@v1
        with:
          prompt: |
            Summarize the following GitHub issue in one paragraph:
            Title: ${{ github.event.issue.title }}
            Body: ${{ github.event.issue.body }}

      - name: Comment with AI summary
        run: |
          gh issue comment $ISSUE_NUMBER --body '${{ steps.inference.outputs.response }}'
        env:
          GH_TOKEN: ${{ secrets.GITHUB_TOKEN }}
          ISSUE_NUMBER: ${{ github.event.issue.number }}
          RESPONSE: ${{ steps.inference.outputs.response }}
\end{lstlisting}

%% file: reference.bib
@misc{github_actions,
  author       = {{GitHub}},
  title        = {Understanding GitHub Actions},
    year         = {2026},
  howpublished = {\url{https://docs.github.com/en/actions/get-started/understand-github-actions}},
  note         = {Accessed: 2026-05-06}
}

@misc{github_agentic_workflows_2026,
  author       = {{GitHub}},
  title        = {GitHub Agentic Workflows},
  year         = {2026},
  url          = {https://github.github.com/gh-aw/},
  note         = {Accessed: 2026-05-06}
}

@misc{anthropic_claude_code_action_2026,
  author       = {{Anthropic}},
  title        = {claude-code-action},
  year         = {2026},
  url          = {https://github.com/anthropics/claude-code-action},
  note         = {Accessed: 2026-05-06}
}

@misc{openai_codex_action_2026,
  author       = {{OpenAI}},
  title        = {codex-action},
  year         = {2026},
  url          = {https://github.com/openai/codex-action},
  note         = {Accessed: 2026-05-06}
}

@misc{google_run_gemini_cli_2026,
  author       = {{Google}},
  title        = {run-gemini-cli},
  year         = {2026},
  url          = {https://github.com/google-github-actions/run-gemini-cli},
  note         = {Accessed: 2026-05-06}
}

@misc{github_marketplace_ai_assisted_actions_2026,
  author       = {{GitHub}},
  title        = {Marketplace: AI Assisted Actions},
  year         = {2026},
  url          = {https://github.com/marketplace?type=actions&category=ai-assisted},
  note         = {Accessed: 2026-05-06}
}

@misc{github_marketplace,
  author       = {{GitHub}},
  title        = {GitHub Action Marketplace},
  year         = {2026},
  url          = {https://github.com/marketplace?type=actions},
  note         = {Accessed: 2026-05-06}
}

@misc{github_action_event_trigger,
  author       = {{GitHub}},
  title        = {Events that trigger workflows},
  year         = {2026},
  url          = {https://docs.github.com/en/actions/reference/workflows-and-actions/events-that-trigger-workflows},
  note         = {Accessed: 2026-05-06}
}

@misc{github_action_syntax,
  author       = {{GitHub}},
  title        = {Workflow syntax for GitHub Actions},
  year         = {2026},
  url          = {https://docs.github.com/en/actions/reference/workflows-and-actions/workflow-syntax},
  note         = {Accessed: 2026-05-06}
}

@misc{github_docs_script_injections,
  author       = {{GitHub Docs}},
  title        = {Script injections},
  howpublished = {\url{https://docs.github.com/en/actions/concepts/security/script-injections}},
  note         = {Accessed: 2026-05-06},
  year         = {2026}
}

@misc{ken_muse_github_actions_injection_attacks,
  author       = {Ken Muse},
  title        = {GitHub Actions Injection Attacks},
  howpublished = {\url{https://www.kenmuse.com/blog/github-actions-injection-attacks/}},
  note         = {Accessed: 2026-05-06},
  year         = {2023}
}

@misc{github_security_lab_preventing_pwn_requests,
  author       = {Jaroslav Lobačevski},
  title        = {Keeping your GitHub Actions and workflows secure Part 1: Preventing pwn requests},
  howpublished = {\url{https://securitylab.github.com/resources/github-actions-preventing-pwn-requests/}},
  note         = {Accessed: 2026-05-06},
  year         = {2021}
}

@misc{github_security_lab_untrusted_input,
  author       = {Jaroslav Lobačevski},
  title        = {Keeping your GitHub Actions and workflows secure Part 2: Untrusted input},
  howpublished = {\url{https://securitylab.github.com/research/github-actions-untrusted-input/}},
  note         = {Accessed: 2026-05-06},
  year         = {2021}
}

@misc{github_docs_variables,
  author       = {{GitHub Docs}},
  title        = {Variables},
  howpublished = {\url{https://docs.github.com/en/actions/concepts/workflows-and-actions/variables}},
  year         = {2026},
  note         = {Accessed: 2026-05-06}
}

@misc{github_docs_contexts,
  author       = {{GitHub Docs}},
  title        = {Contexts},
  howpublished = {\url{https://docs.github.com/en/actions/concepts/workflows-and-actions/contexts}},
  year         = {2026},
  note         = {Accessed: 2026-05-06}
}

@misc{github_docs_expressions,
  author       = {{GitHub Docs}},
  title        = {Expressions},
  howpublished = {\url{https://docs.github.com/en/actions/concepts/workflows-and-actions/expressions}},
  year         = {2026},
  note         = {Accessed: 2026-05-06}
}

@misc{hacktricks_gh_actions_cache_poisoning,
  author       = {{HackTricks Cloud}},
  title        = {{Cache Poisoning}},
  howpublished = {\url{https://cloud.hacktricks.wiki/en/pentesting-ci-cd/github-security/abusing-github-actions/gh-actions-cache-poisoning.html}},
  note         = {Accessed: 2026-05-06}
}

@misc{hacktricks_gh_actions_artifact_poisoning,
  author       = {{HackTricks Cloud}},
  title        = {{Artifact Poisoning}},
  howpublished = {\url{https://cloud.hacktricks.wiki/en/pentesting-ci-cd/github-security/abusing-github-actions/gh-actions-artifact-poisoning.html}},
  note         = {Accessed: 2026-05-06}
}

@misc{hacktricks_gh_actions_context_script_injections,
  author       = {{HackTricks Cloud}},
  title        = {{Context Script Injections}},
  howpublished = {\url{https://cloud.hacktricks.wiki/en/pentesting-ci-cd/github-security/abusing-github-actions/gh-actions-context-script-injections.html}},
  note         = {Accessed: 2026-05-06}
}

@misc{openssf_mitigating_attack_vectors_github_workflows,
  author       = {{Open Source Security Foundation}},
  title        = {Mitigating Attack Vectors in GitHub Workflows},
  howpublished = {\url{https://openssf.org/blog/2024/08/12/mitigating-attack-vectors-in-github-workflows/}},
  note         = {Accessed: 2026-05-06},
  year         = {2024}
}

@inproceedings{liu2025agentfuzz,
author = {Liu, Fengyu and Zhang, Yuan and Luo, Jiaqi and Dai, Jiarun and Chen, Tian and Yuan, Letian and Yu, Zhengmin and Shi, Youkun and Li, Ke and Zhou, Chengyuan and Chen, Hao and Yang, Min},
title = {Make agent defeat agent: automatic detection of taint-style vulnerabilities in LLM-based agents},
year = {2025},
isbn = {978-1-939133-52-6},
publisher = {USENIX Association},
address = {USA},
booktitle = {Proceedings of the 34th USENIX Conference on Security Symposium},
articleno = {194},
numpages = {19},
location = {Seattle, WA, USA},
series = {SEC '25}
}

@inproceedings{liu2024llmrce,
author = {Liu, Tong and Deng, Zizhuang and Meng, Guozhu and Li, Yuekang and Chen, Kai},
title = {Demystifying RCE Vulnerabilities in LLM-Integrated Apps},
year = {2024},
isbn = {9798400706363},
publisher = {Association for Computing Machinery},
address = {New York, NY, USA},
url = {https://doi.org/10.1145/3658644.3690338},
doi = {10.1145/3658644.3690338},
booktitle = {Proceedings of the 2024 on ACM SIGSAC Conference on Computer and Communications Security},
pages = {1716–1730},
numpages = {15},
location = {Salt Lake City, UT, USA},
series = {CCS '24}
}

@inproceedings{icse26taintp2x,
author = {He, Junjie and Wang, Shenao and Zhao, Yanjie and Hou, Xinyi and Liu, Zhao and Zou, Quanchen and Wang, Haoyu},
title = {TaintP2X: Detecting Taint-Style Prompt-to-Anything Injection Vulnerabilities in LLM-Integrated Applications},
year = {2026},
booktitle = {Proceedings of the IEEE/ACM 48th International Conference on Software Engineering},
doi = {10.1145/3744916.3773199},
url = {https://doi.org/10.1145/3744916.3773199}
}

@INPROCEEDINGS {pedro2025prompt2sql,
author = { Pedro, Rodrigo and E. Coimbra, Miguel and Castro, Daniel and Carreira, Paulo and Santos, Nuno },
booktitle = { 2025 IEEE/ACM 47th International Conference on Software Engineering (ICSE) },
title = {{ Prompt-to-SQL Injections in LLM-Integrated Web Applications: Risks and Defenses }},
year = {2025},
volume = {},
ISSN = {1558-1225},
pages = {76-88},
doi = {10.1109/ICSE55347.2025.00007},
url = {https://doi.ieeecomputersociety.org/10.1109/ICSE55347.2025.00007},
publisher = {IEEE Computer Society},
address = {Los Alamitos, CA, USA},
month =May}

@article{wu2026chainfuzzer,
  author       = {Jiangrong Wu and
                  Zitong Yao and
                  Yuhong Nan and
                  Zibin Zheng},
  title        = {ChainFuzzer: Greybox Fuzzing for Workflow-Level Multi-Tool Vulnerabilities
                  in {LLM} Agents},
  journal      = {CoRR},
  volume       = {abs/2603.12614},
  year         = {2026},
  url          = {https://doi.org/10.48550/arXiv.2603.12614},
  doi          = {10.48550/ARXIV.2603.12614},
  eprinttype   = {arXiv},
  eprint       = {2603.12614},
  bibsource    = {dblp computer science bibliography, https://dblp.org}
}

@misc{aikido_promptpwnd,
  author       = {Rein Daelman},
  title        = {PromptPwnd: Prompt Injection Vulnerabilities in GitHub Actions Using AI Agents},
  howpublished = {\url{https://www.aikido.dev/blog/promptpwnd-github-actions-ai-agents}},
  note         = {Accessed: 2026-05-06},
  year         = {2025}
}

@misc{snyk_clinejection,
  author       = {Stephen Thoemmes},
  title        = {How ``Clinejection'' Turned an AI Bot into a Supply Chain Attack},
  howpublished = {\url{https://snyk.io/blog/cline-supply-chain-attack-prompt-injection-github-actions/}},
  note         = {Accessed: 2026-05-06},
  year         = {2026}
}

@article{shi2025promptinjection,
  author       = {Jiawen Shi and
                  Zenghui Yuan and
                  Guiyao Tie and
                  Pan Zhou and
                  Neil Zhenqiang Gong and
                  Lichao Sun},
  title        = {Prompt Injection Attack to Tool Selection in {LLM} Agents},
  journal      = {CoRR},
  volume       = {abs/2504.19793},
  year         = {2025},
  url          = {https://doi.org/10.48550/arXiv.2504.19793},
  doi          = {10.48550/ARXIV.2504.19793},
  eprinttype   = {arXiv},
  eprint       = {2504.19793},
  bibsource    = {dblp computer science bibliography, https://dblp.org}
}

@misc{li2025crosstoolharvesting,
      title={Les Dissonances: Cross-Tool Harvesting and Polluting in Pool-of-Tools Empowered LLM Agents}, 
      author={Zichuan Li and Jian Cui and Xiaojing Liao and Luyi Xing},
      year={2025},
      eprint={2504.03111},
      archivePrefix={arXiv},
      primaryClass={cs.CR},
      url={https://arxiv.org/abs/2504.03111}, 
}

@inproceedings{muralee23argus,
author = {Muralee, Siddharth and Koishybayev, Igibek and Nahapetyan, Aleksandr and Tystahl, Greg and Reaves, Brad and Bianchi, Antonio and Enck, William and Kapravelos, Alexandros and Machiry, Aravind},
title = {ARGUS: a framework for staged static taint analysis of GitHub workflows and actions},
year = {2023},
isbn = {978-1-939133-37-3},
publisher = {USENIX Association},
address = {USA},
booktitle = {Proceedings of the 32nd USENIX Conference on Security Symposium},
articleno = {391},
numpages = {18},
location = {Anaheim, CA, USA},
series = {SEC '23}
}

@article{madjda2026actionscanner,
  author       = {Madjda Fares and
                  Yogya Gamage and
                  Benoit Baudry},
  title        = {Unpacking Security Scanners for GitHub Actions Workflows},
  journal      = {CoRR},
  volume       = {abs/2601.14455},
  year         = {2026},
  url          = {https://doi.org/10.48550/arXiv.2601.14455},
  doi          = {10.48550/ARXIV.2601.14455},
  eprinttype   = {arXiv},
  eprint       = {2601.14455},
  bibsource    = {dblp computer science bibliography, https://dblp.org}
}

@inproceedings{igibek22characterizing,
  author       = {Igibek Koishybayev and
                  Aleksandr Nahapetyan and
                  Raima Zachariah and
                  Siddharth Muralee and
                  Bradley Reaves and
                  Alexandros Kapravelos and
                  Aravind Machiry},
  editor       = {Kevin R. B. Butler and
                  Kurt Thomas},
  title        = {Characterizing the Security of Github {CI} Workflows},
  booktitle    = {31st {USENIX} Security Symposium, {USENIX} Security 2022, Boston,
                  MA, USA, August 10-12, 2022},
  pages        = {2747--2763},
  publisher    = {{USENIX} Association},
  year         = {2022},
  url          = {https://www.usenix.org/conference/usenixsecurity22/presentation/koishybayev},
  bibsource    = {dblp computer science bibliography, https://dblp.org}
}

@INPROCEEDINGS{huang2025revisiting,
  author={Huang, Jiangnan and Lin, Bin},
  booktitle={2025 IEEE/ACM 33rd International Conference on Program Comprehension (ICPC)}, 
  title={Revisiting Security Practices for Github Actions Workflows}, 
  year={2025},
  volume={},
  number={},
  pages={73-77},
  doi={10.1109/ICPC66645.2025.00016}
}

@inproceedings{benedetti22assessment, 
   title={Automatic Security Assessment of GitHub Actions Workflows},
   url={http://dx.doi.org/10.1145/3560835.3564554},
   DOI={10.1145/3560835.3564554},
   booktitle={Proceedings of the 2022 ACM Workshop on Software Supply Chain Offensive Research and Ecosystem Defenses},
   publisher={ACM},
   author={Benedetti, Giacomo and Verderame, Luca and Merlo, Alessio},
   year={2022},
   month=nov, 
   pages={37–45}
}

@article{zheng2026actionfail,
author = {Zheng, Lianyu and Li, Shuang and Huang, Xi and Huang, Jiangnan and Lin, Bin and Chen, Jinfu and Xuan, Jifeng},
title = {Why Do GitHub Actions Workflows Fail? An Empirical Study},
year = {2026},
issue_date = {May 2026},
publisher = {Association for Computing Machinery},
address = {New York, NY, USA},
volume = {35},
number = {5},
issn = {1049-331X},
url = {https://doi.org/10.1145/3749371},
doi = {10.1145/3749371},
journal = {ACM Trans. Softw. Eng. Methodol.},
month = apr,
articleno = {139},
numpages = {29}
}

@misc{prompt_injection_scanner,
  title        = {{Prompt Injection Scanner}},
  author       = {{alexh-scrt}},
  howpublished = {\url{https://github.com/alexh-scrt/prompt-injection-scanner}},
  note         = {Accessed: 2026-05-06}
}

@misc{aikido_opengrep_rules,
  title        = {{Aikido OpenGrep Rules}},
  author       = {{Aikido Security}},
  howpublished = {\url{https://github.com/AikidoSec/opengrep-rules}},
  note         = {Accessed: 2026-05-06}
}

@misc{trailofbits_agentic_actions_auditor,
  title        = {{Agentic Actions Auditor}},
  author       = {{Trail of Bits}},
  howpublished = {\url{https://github.com/trailofbits/skills/tree/main/plugins/agentic-actions-auditor}},
  note         = {Accessed: 2026-05-06}
}

@misc{codeql_actions_queries,
  title        = {{CodeQL Actions Query Pack}},
  author       = {{GitHub}},
  howpublished = {\url{https://github.com/github/codeql/tree/main/actions/ql}},
  note         = {Accessed: 2026-05-06}
}

@article{wang2026yasa,
  author       = {Yayi Wang and
                  Shenao Wang and
                  Jian Zhao and
                  Shaosen Shi and
                  Ting Li and
                  Yan Cheng and
                  Lizhong Bian and
                  Kan Yu and
                  Yanjie Zhao and
                  Haoyu Wang},
  title        = {{YASA:} Scalable Multi-Language Taint Analysis on the Unified {AST}
                  at Ant Group},
  journal      = {CoRR},
  volume       = {abs/2601.17390},
  year         = {2026},
  url          = {https://doi.org/10.48550/arXiv.2601.17390},
  doi          = {10.48550/ARXIV.2601.17390},
  eprinttype   = {arXiv},
  eprint       = {2601.17390},
  bibsource    = {dblp computer science bibliography, https://dblp.org}
}

@article{weber2024llmapp,
  author       = {Irene Weber},
  title        = {Large Language Models as Software Components: {A} Taxonomy for LLM-Integrated
                  Applications},
  journal      = {CoRR},
  volume       = {abs/2406.10300},
  year         = {2024},
  url          = {https://doi.org/10.48550/arXiv.2406.10300},
  doi          = {10.48550/ARXIV.2406.10300},
  eprinttype    = {arXiv},
  eprint       = {2406.10300},
  bibsource    = {dblp computer science bibliography, https://dblp.org}
}

@article{wang2024agent,
  author       = {Lei Wang and
                  Chen Ma and
                  Xueyang Feng and
                  Zeyu Zhang and
                  Hao Yang and
                  Jingsen Zhang and
                  Zhiyuan Chen and
                  Jiakai Tang and
                  Xu Chen and
                  Yankai Lin and
                  Wayne Xin Zhao and
                  Zhewei Wei and
                  Jirong Wen},
  title        = {A survey on large language model based autonomous agents},
  journal      = {Frontiers Comput. Sci.},
  volume       = {18},
  number       = {6},
  pages        = {186345},
  year         = {2024},
  url          = {https://doi.org/10.1007/s11704-024-40231-1},
  doi          = {10.1007/S11704-024-40231-1},
  bibsource    = {dblp computer science bibliography, https://dblp.org}
}

@article{he2024emerged,
  author       = {Feng He and
                  Tianqing Zhu and
                  Dayong Ye and
                  Bo Liu and
                  Wanlei Zhou and
                  Philip S. Yu},
  title        = {The Emerged Security and Privacy of {LLM} Agent: {A} Survey with Case
                  Studies},
  journal      = {CoRR},
  volume       = {abs/2407.19354},
  year         = {2024},
  url          = {https://doi.org/10.48550/arXiv.2407.19354},
  doi          = {10.48550/ARXIV.2407.19354},
  eprinttype    = {arXiv},
  eprint       = {2407.19354},
  bibsource    = {dblp computer science bibliography, https://dblp.org}
}

@article{deng2025agents,
author = {Deng, Zehang and Guo, Yongjian and Han, Changzhou and Ma, Wanlun and Xiong, Junwu and Wen, Sheng and Xiang, Yang},
title = {AI Agents Under Threat: A Survey of Key Security Challenges and Future Pathways},
year = {2025},
issue_date = {July 2025},
publisher = {Association for Computing Machinery},
address = {New York, NY, USA},
volume = {57},
number = {7},
issn = {0360-0300},
url = {https://doi.org/10.1145/3716628},
doi = {10.1145/3716628},
journal = {ACM Comput. Surv.},
month = feb,
articleno = {182},
numpages = {36}
}

@article{zhang2024agent,
  author       = {Hanrong Zhang and
                  Jingyuan Huang and
                  Kai Mei and
                  Yifei Yao and
                  Zhenting Wang and
                  Chenlu Zhan and
                  Hongwei Wang and
                  Yongfeng Zhang},
  title        = {Agent Security Bench {(ASB):} Formalizing and Benchmarking Attacks
                  and Defenses in LLM-based Agents},
  journal      = {CoRR},
  volume       = {abs/2410.02644},
  year         = {2024},
  url          = {https://doi.org/10.48550/arXiv.2410.02644},
  doi          = {10.48550/ARXIV.2410.02644},
  eprinttype    = {arXiv},
  eprint       = {2410.02644},
  bibsource    = {dblp computer science bibliography, https://dblp.org}
}

@inproceedings{luo2026agentauditor,
title={AgentAuditor: Human-level Safety and Security Evaluation for {LLM} Agents},
author={Hanjun Luo and Shenyu Dai and Chiming Ni and Xinfeng Li and Guibin Zhang and Kun Wang and Tongliang Liu and Hanan Salam},
booktitle={The Thirty-ninth Annual Conference on Neural Information Processing Systems},
year={2026},
url={https://openreview.net/forum?id=2KKqp7MWJM}
}

@misc{anthropic_claude_code_action_security,
  author       = {{Anthropic}},
  title        = {{Claude Code Action Security Policy}},
  howpublished = {\url{https://github.com/anthropics/claude-code-action/blob/main/docs/security.md}},
  note         = {Accessed: 2026-05-06}
}

@misc{openai_codex_action_security,
  author       = {{OpenAI}},
  title        = {{Codex Action Security Policy}},
  howpublished = {\url{https://github.com/openai/codex-action/blob/main/docs/security.md}},
  note         = {Accessed: 2026-05-06}
}
